\documentclass[acmtog]{acmart}  
\usepackage{mysiggraph}

\setcopyright{acmlicensed}
\acmJournal{TOG}
\acmYear{2020}
\acmVolume{39}
\acmNumber{6}
\acmArticle{164}
\acmMonth{12} 
\acmDOI{10.1145/3414685.3417807}

\begin{document}
\title{\name: Sequential CAD Modeling by Sketching in Context}

\author{Changjian Li}
\affiliation{
 \institution{University College London}
 \streetaddress{66-72 Gower Street}
 \city{London}
 }
\email{changjian.li@ucl.ac.uk}

\author{Hao Pan}
\affiliation{
 \institution{Microsoft Research Asia}
 \streetaddress{No.5 Danling Rd}
 \city{Beijing}
}
\email{haopan@microsoft.com}

\author{Adrien Bousseau}
\affiliation{
 \institution{Inria, Universit\'{e} C\^{o}te d'Azur}
 \streetaddress{2004 route des lucioles}
 \city{Valbonne}
}
\email{adrien.bousseau@inria.fr}

\author{Niloy J. Mitra}
\affiliation{
\institution{University College London}
\streetaddress{66-72 Gower Street}
\city{London}
}
\affiliation{
\institution{Adobe Research}
}
\email{n.mitra@cs.ucl.ac.uk}


\begin{teaserfigure}
\centering
\begin{overpic}[width=\linewidth]{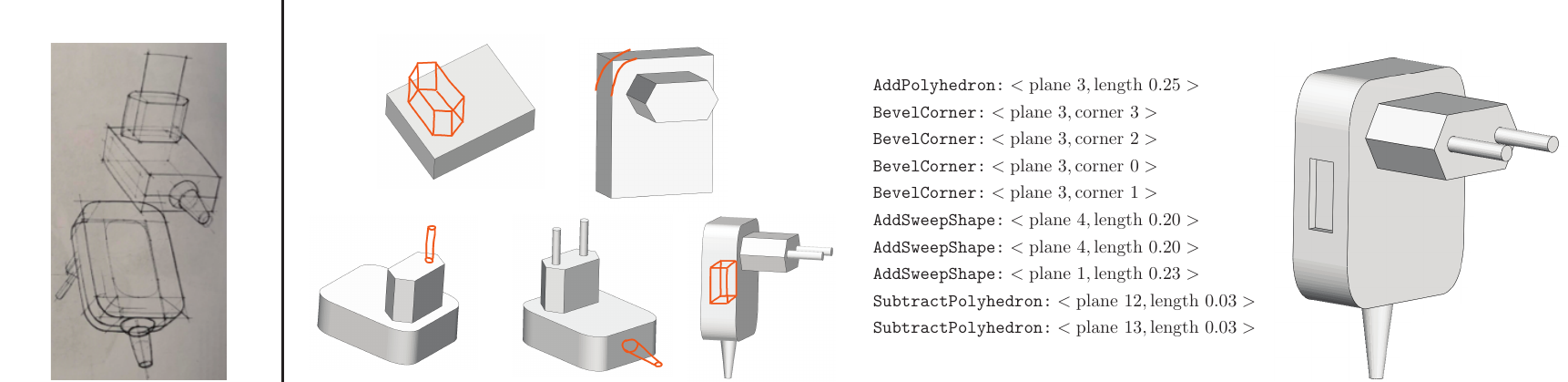}
\put(-1, 23) {\bf{(a) inspirational sketch}}
\put(19, 23) {\bf{(b) sketching sequence using Sketch2CAD}}
\put(55, 23) {\bf{(c) inferred CAD instructions}}
\put(82, 23) {\bf{(d) CAD model}}
\put(25,12) {\small step 1}
\put(41, 12) {\small step 2}
\put(20, 1) {\small step 6}
\put(31, 1) {\small step 8}
\put(48, 1) {\small step 9}
\end{overpic}
\caption{Industrial designers commonly decompose complex shapes into box-like primitives, which they refine by drawing cuts and roundings, or by adding and substracting smaller parts~\cite{eissen2007sketching,eissen2011sketching}~(a, \rev{$\copyright$Koos Eissen and
Roselien Steur}). Users of \emph{Sketch2CAD} follow similar sketching steps (b), which our system interprets as parametric modeling operations (c) to automatically output a precise, compact, and editable CAD model (d).}
\label{fig:teaser}
\end{teaserfigure}
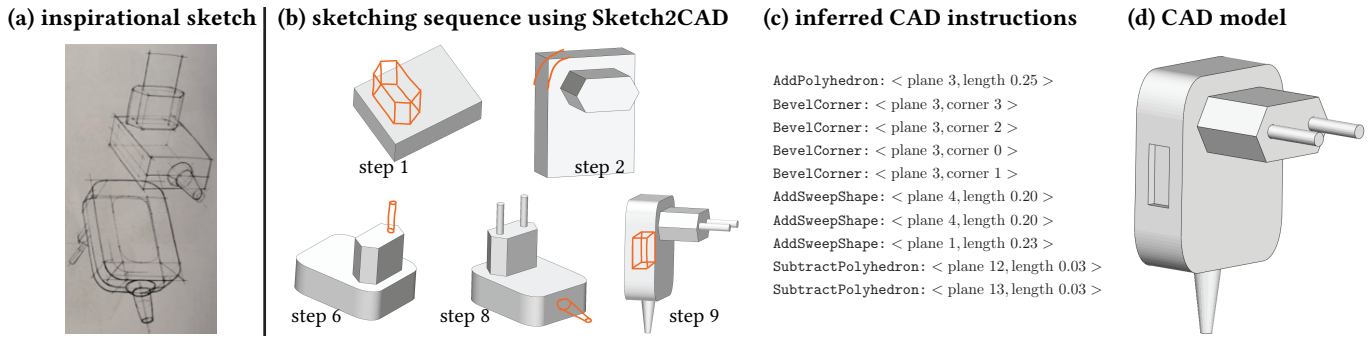

\begin{abstract}
We present a sketch-based CAD modeling system, where users create objects incrementally by sketching the desired shape edits, which our system automatically translates to CAD operations. Our approach is motivated by the close similarities between the steps industrial designers follow to draw 3D shapes, and the operations CAD modeling systems offer to create similar shapes. To overcome the strong ambiguity with parsing 2D sketches, we observe that in a sketching sequence, each step makes sense and can be interpreted in the \emph{context} of what has been drawn before. In our system, this context corresponds to a partial CAD model, inferred in the previous steps, which we feed along with the input sketch to a deep neural network in charge of interpreting how the model should be modified by that sketch. Our deep network architecture then recognizes the intended CAD operation and segments the sketch accordingly, such that a subsequent optimization estimates the parameters of the operation that best fit the segmented sketch strokes. Since there exists no datasets of paired sketching and CAD modeling sequences, we train our system by generating synthetic sequences of CAD operations that we render as line drawings. We present a proof of concept realization of our algorithm supporting four frequently used CAD operations. Using our system,  participants are able to quickly model a large and diverse set of objects,  demonstrating \name to be an  alternate way of interacting with current CAD modeling systems.
\end{abstract}

%
%



\maketitle

\begin{figure*}[t!]
    \centering
    \begin{overpic}[width=\linewidth]{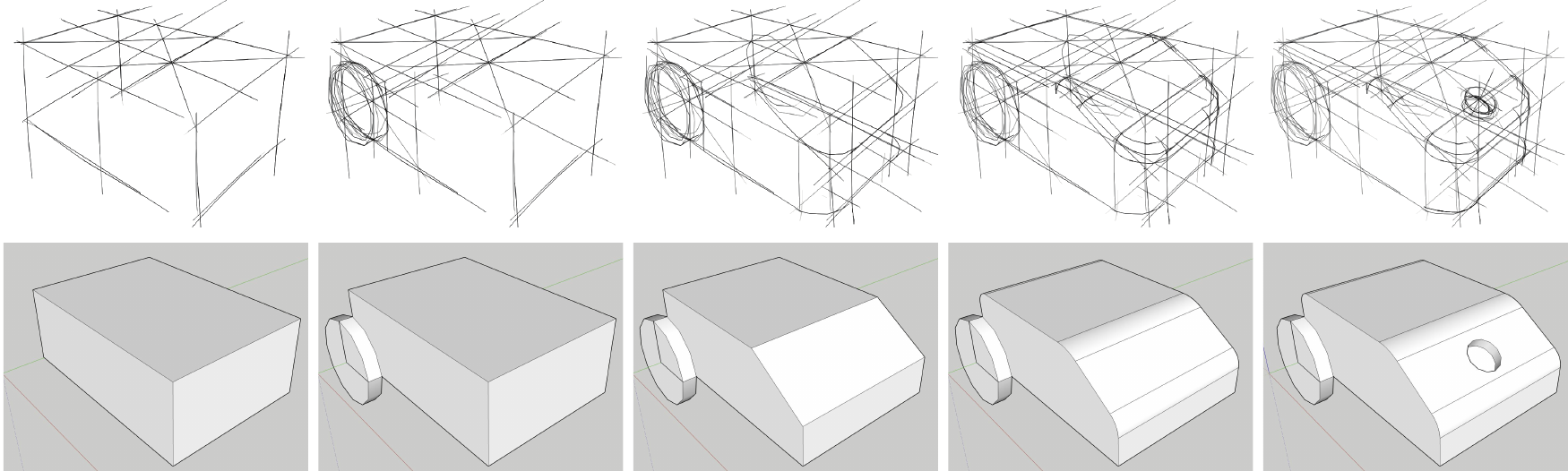}
    \put(0.5, 29) {\small (a)}
    \put(0.5, 12.5) {\small (b)}
    \put(0.5, 15) {\small create box}
    \put(21, 15) {\small add cylinder}
    \put(41, 15) {\small bevel box}
    \put(61, 15) {\small smooth edge}
    \put(81, 15) {\small subtract cylinder}
    \end{overpic}
    \caption{\textbf{Similarity between sketching and CAD modeling workflows.} (a) As illustrated by this sequence from the \emph{OpenSketch} dataset \protect\cite{GSHPDB19}, industrial designers construct their drawings by starting from a simple shape (here a box) that they refine by adding or subtracting sub-parts (wheels, beveled edges, hole). (b) Modern CAD software such as SketchUP \protect\cite{sketchup} rely on very similar operations to model 3D shapes. }
    \label{fig:motivation}
    \vspace{-4mm}
\end{figure*}
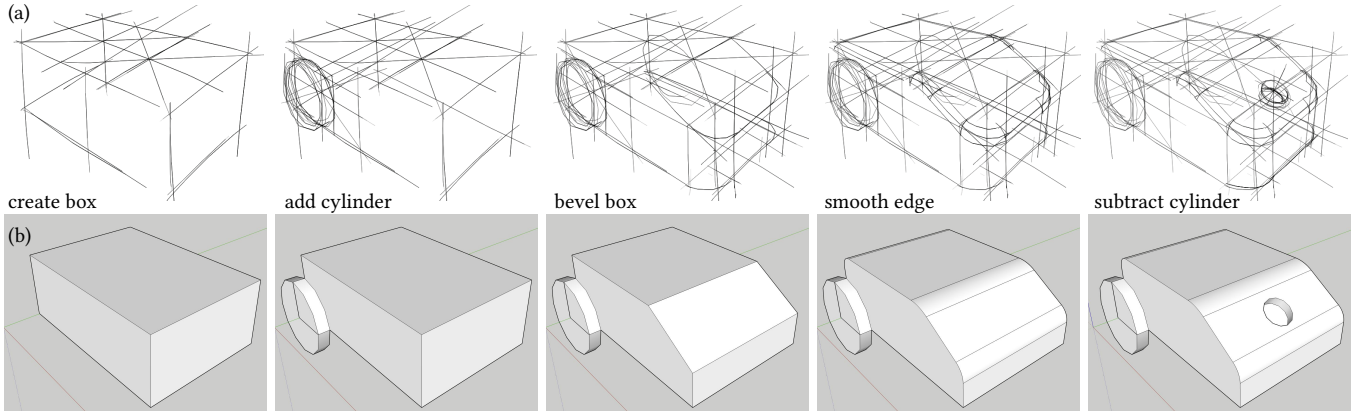

\section{Introduction}
Sketching and 3D modeling are two major steps of industrial design. Sketching is typically done first, as it allows designers to express their vision quickly and approximately \cite{eissen2007sketching}. Design sketches are then converted to 3D models for downstream engineering and manufacturing, using CAD tools that offer high precision and editability \cite{pipes2007drawing}. However, design sketching and CAD modeling are often performed by different experts with different skill sets, making design iterations cumbersome, expensive, and time consuming.

While a number of methods have been proposed to create 3D models by sketching, existing solutions often lack the precision and editability of CAD modeling. On the one hand, interactive systems interpret user strokes as custom modeling operations rather than generic CAD \cite{igarashiTeddy,bae2008ilovesketch,Nishida:2016:InverseProcedural,Zeleznik:1996:SAI}. On the other hand, methods that interpret complete sketches are limited to specific drawing techniques \cite{xu2014true2form} and classes of shapes \cite{lun20173d}, and output curve networks or triangular meshes rather than editable models. As stressed by a recent survey \cite{bonnici2019sketch}, to be widely used by the design industry, \textit{``sketch-based modeling systems should integrate seamlessly with existing workflow practices''}. 

Our key observation is that despite their apparent differences, design sketching and CAD modeling actually involve very similar workflows, yet expressed in different languages. Industrial designers often start their sketches by drawing the overall shape as an assemblage of boxes and cylinders called \emph{scaffolds}, which they then refine by drawing roundings, sub-parts, and small details (Fig.~\ref{fig:motivation}a). Similarly, CAD modelers often start with simple geometric primitives that they refine to build up complex models by progressively applying geometric operations (e.g., extrude, bevel, smooth) (Fig.~\ref{fig:motivation}b).
Based on this observation, we propose \name as a learning-based interactive modeling system that translates sketching operations into their corresponding CAD modeling operations. Users of our system thus express their ideas using similar sketching steps as they would do on paper, yet obtain as output a regular CAD model, along with a trace of the sequential operations, ready to be fabricated or further edited with existing CAD software. Our system can be seen as a translator that interprets user drawn strokes in context of the current modeling session, and maps them into a sequence of predefined CAD operations, along with their associated parameters. The system empowers users to create regular CAD models without having to navigate complex CAD system interfaces. 

We first propose a common parameterization of popular sketching and CAD modeling operations (e.g., \texttt{extrude}, \texttt{bevel}, \texttt{add}, \texttt{subtract}, \texttt{sweep}). For each operation, our parameterization encodes the different components of the CAD shape, which correspond to different strokes in the sketch. For instance, a \texttt{bevel} operation is composed of two parallel curves that define the new profile of the corner on which it applies. Importantly, our parameterization also encodes the faces of the current 3D model that should be modified by the operation, since CAD operations are typically applied in sequence to progressively achieve complex shapes.

The main challenge is then to recover, for every step of a sketching session, the intended modeling operation and the asociated parameters. This is a highly ambiguous task, not only because the strokes are often imprecise, but also because similar strokes might have different meaning depending on the \emph{context} in which they are drawn. 
We propose a three-stage pipeline that progressively reduces this ambiguity to produce  regular CAD objects. 
The first stage \emph{classifies} the sketch among possible CAD operations (\texttt{extrude}, \texttt{bevel}, \texttt{add}, \texttt{subtract}, \texttt{sweep}). In addition to the user sketch, the classifier takes as input depth and normal maps of the current 3D model, which provides strong \emph{contextual cues} about the intended operation. The second stage \emph{segments} the user sketch and contextual maps into parts, specific to the target CAD operation. For instance, the sketch of a \texttt{bevel} operation is segmented into its two profile curves, while the contextual maps are segmented to form a mask of the face on which the \texttt{bevel} operation needs to be applied. Finally, the third stage \emph{instantiates} the CAD operation by fitting parametric curves or shapes on the segmented strokes and projecting these strokes, optionally regularized, on the selected faces of the 3D model.

From a technical standpoint, we realize our classification and segmentation stages with deep convolutional networks. In addition to the design of CAD-specific segmentation networks, a contribution of our work resides in a large training dataset of CAD-like objects that we synthetically generated by sampling sequences of CAD operations. We took special care in balancing this dataset such that the most complex operations appear more frequently, and that parameters of all operations are sampled uniformly. Furthermore, we also balanced the length of the operation sequences, such that our system can recognize CAD operations at any stage of a modeling session.  

In contrast to prior learning-based methods that were trained on particular domains \cite{Nishida:2016:InverseProcedural,huang2016shape} or selected object classes \cite{lun20173d}, a key strength of our approach is that it recognizes \rev{a set of existing} CAD operations that can be applied in arbitrary order, allowing the creation of a \rev{diverse} range of human-made objects. In addition, the parametric nature of each such operation results in shapes that are highly precise and regular despite very approximate input strokes. Figure~\ref{fig:teaser} shows a typical modeling session using \name. Finally, since our training is entirely synthetic, \rev{we believe that the same approach can be used to} extend to support other operations.

\rev{While our sketch-based modeling system does not provide the same level of comprehensive modeling as modern CAD software like SketchUP \cite{sketchup} and TinkerCAD \cite{tinkercad}, it demonstrates an alternate way of interacting with existing CAD systems without requiring repeated command selection and switching. Our interface can be particularly attractive to product designers or novice users who are more fluent with sketching than with CAD modeling interfaces. By allowing non-experts to quickly produce complete CAD protocols (see Sec.~\ref{sec:user_study}), our tool holds the potential to facilitate more direct collaboration between novices and experts.}

In summary, our main contributions are:
\begin{itemize}
\item formalizing a set of common CAD operations and their corresponding sketches, allowing an automatic translation between the two domains; 
\item developing a pipeline of deep neural networks capable of recognizing and segmenting CAD operations from sketches drawn over 3D shapes, and producing precise, regular 3D geometry by fitting CAD parameters on the predictions; 
\item designing a large dataset of synthetic CAD models, along with their step-by-step construction sequences; and, as a culmination of these taken together,  
\item presenting \name as a novel sketch-based modeling system that unifies the sequential workflows of product design sketching and CAD modeling.
\end{itemize}

Code, training data, and the \name system are available on the project page for research use.

\section{Related Work}
Our work aims to bridge the gap between sketch-based and CAD modeling.

\paragraph{CAD modeling.}
Computer-Aided Design has long been adopted by the industry to create precise and high-quality 3D models suitable for physical simulation, lighting simulation, and downstream manufacturing \cite{3dsmax,maya,rhino,sketchup}. However, the high precision offered by CAD comes at the price of complex interfaces to allow users to select appropriate geometric operations and tune their parameters. Various approaches have been considered to reduce this user burden, from automatic alignment of existing CAD models on scanned point clouds \cite{avetisyan2019scan2cad}, to educational visualizations of modeling sequences \cite{meshflow2011}. We contribute to this effort by   instantiating CAD operations by sequentially interpreting hand-drawn sketches.

Closer to our work are methods aiming at converting raw 3D meshes into editable CAD models, which can be formulated as a form of program synthesis \cite{Sharma_2018_CVPR,TaoInverseCSG,tian2018learning}. On the one hand, leveraging the sequential nature of sketching and CAD modeling makes our problem better posed than the conversion of complete objects that these methods target. On the other hand, we take as input approximate sketch lines rather than precise 3D models, which induces additional ambiguity. \rev{Ellis et al.~\shortcite{ellis2018learning} also applied program synthesis to convert sketches to graphics programs, but focused on 2D diagrams and as such did not consider depth recovery.}

\paragraph{Sketch-based modeling.}
Existing work on sketch-based modeling can be broadly classified into \emph{offline} and \emph{online} methods.
Offline methods aim at interpreting complete drawings, either automatically or with user assistance. 
Early algorithms detect geometric constraints between curves, such as parallelism, orthogonality and symmetry, and solve for the 3D curve network that best satisfies these constraints \cite{lipson1996optimization,naya2002direct,Cordier2013,xu2014true2form,Wang09}.
The main limitation of these methods is that they require clean drawings as input to detect and enforce relevant constraints. In addition, the curve networks they produce are not directly usable by downstream 3D modeling and simulation software. These limitations are partly addressed by interactive tools that allow users to align geometric primitives over the drawing \cite{Gingold09,Shtof2013}. While the parametric nature of these primitives brings robustness to approximate inputs, users of these systems need to provide a number of annotations to achieve precise alignment and relative positioning. We differ from the above methods by exploiting the common sequential nature of sketching and CAD modeling, which allows us to automatically recognize parametric CAD operations as soon as they are drawn rather than during a subsequent annotation process.

In contrast to the above optimization-based approaches, recent work has explored the potential of deep learning to automatically reconstruct 3D objects from one or several sketches \cite{lun20173d,Su2018,delanoy20183d,li2018robust}. Since these methods build strong shape priors from training data, they are limited to specific classes of objects \cite{lun20173d,Su2018,delanoy20183d} or types of surfaces \cite{li2018robust}. Besides, all these methods predict depth maps or voxel grids that are then converted to triangular meshes, which, in contrast to CAD models, greatly limits the precision and editability of the resulting 3D models. While we also build on powerful deep convolutional networks, a strength of our approach is to predict CAD operations rather than complete objects. Combining these operations allows users to produce a wide variety of parametric shapes, which are precise and editable by construction, and allow for generalization across different CAD models.

\begin{figure*}[t!]
    \centering
    \includegraphics[width=\linewidth]{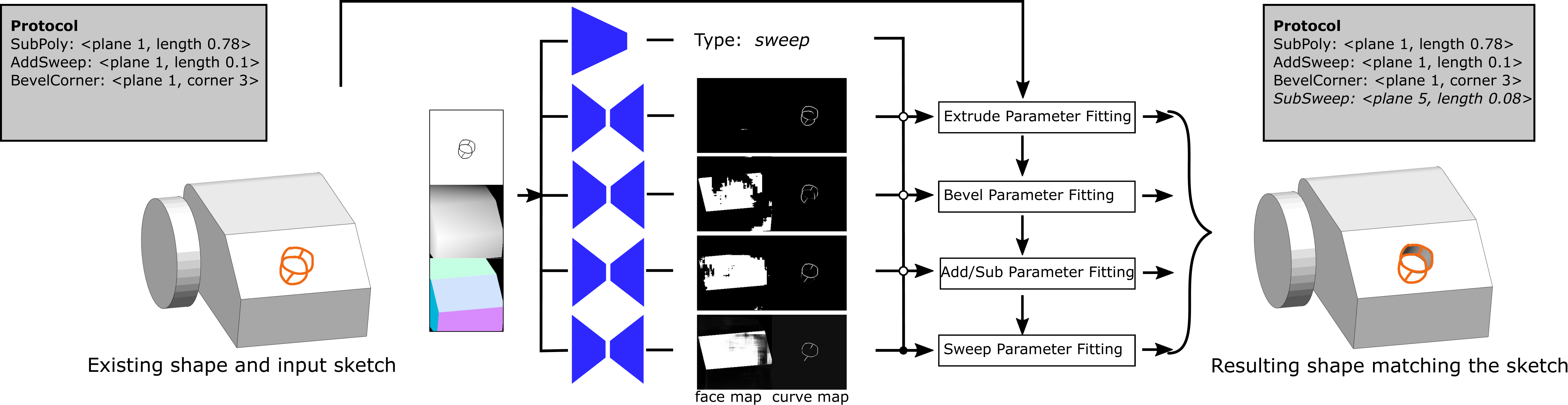}
    \vspace{-6mm}
    \caption{{\bf \name at inference time}. Given an existing shape and input sketch strokes (shown in orange) for the current operation, we first obtain the maps of sketch and local context (i.e., depth and normal), which are fed to the operator classification and segmentation networks. The classified operator type, \texttt{sweep} in this example, is used to select the output base face and curve segmentation maps, based on which the parameters defining the operator are fitted, via an optimization, to recover the sketched operation instance. The recovered operator is then applied to the existing shape to produce the updated model; meanwhile, the operation is pushed into the protocol list.}
    \label{fig:test_stage_pipeline}
    \vspace{-4mm}
\end{figure*}

The sequential workflow we offer makes our method closer in spirit to \emph{online} sketch-based modeling systems, where users create complex 3D shapes incrementally by alternating between 2D sketching and 3D navigation. Because of the difficulty of recovering 3D shapes from 2D strokes, a number of systems focus on specific modeling operations, such as inflation of smooth shapes \cite{igarashiTeddy,Nealen07} or creation of sparse networks of 3D curves \cite{bae2008ilovesketch,schmidt2009analytic}. Closer to our work are methods that enable the creation of CAD models representing man-made shapes. Rivers et al.~\shortcite{Rivers2010} resolve 2D-to-3D ambiguity by asking users to draw the shape parts in three orthographic views, as common in CAD software. We instead let users draw in a single perspective view, as common in product design sketching. \rev{The seminal \emph{SKETCH} system by Zeleznik et al.~\shortcite{Zeleznik:1996:SAI} and \emph{GiDES++} by Jorge et al.~\shortcite{jorge2003gides} include some of our CAD operations. However, users of \emph{SKETCH} specify these operations using a custom vocabulary of sketching \emph{gestures}, while users of \emph{GiDES++} need to decompose object parts into individual strokes interpreted one by one using a set of hand-crafted rules.}
Our originality is to automatically recognize the CAD operations and recover their parameters from freehand sketches, which allows users to directly draw \rev{complete parts of} the shapes they wish to obtain, without requiring to learn a set of new gestures. 
We achieve this recognition using deep neural networks trained on synthetic CAD modeling sequences. 
While Huang et al.~\shortcite{huang2016shape} and Nishida et al.~\shortcite{Nishida:2016:InverseProcedural} explored a similar usage of deep learning for procedural modeling, they target shapes with fixed numbers of parameters, created using a fixed order of operations. For example, Nishida et al. assume that to create a building, users start by sketching the building mass, then the roof, the facades, and finally the windows. In contrast, a major challenge we face is to recognize generic CAD operations with varying number of parameters, sketched in any order. We achieve this goal by accounting for the \emph{context} under which the sketch is drawn. Furthermore, while Huang et al. and Nishida et al. use a regression network to predict the parameters of their shapes, we found this strategy to fail on the more ambiguous problem we target, and instead use a classification network to segment the sketched strokes into CAD-specific components on which we subsequently fit geometric primitives.

We draw inspiration from several earlier systems that explored the possibility to sketch novel shapes in the context of an existing scene, represented as photographs or 3D models \cite{Insitu,Lau2010,FLB15,DePaoli2015,Zheng2016,3DSkectch18,Li2017}. However, these method use the existing context to either guide user sketching or to deduce geometric constraints for lifting the sketch to 3D. Our originality is to leverage context within a sequential modeling workflow, where the existing scene informs the recognition of the intended CAD operation, which aims at modifying that scene.

\section{Method overview}

Suppose we work with solid CAD models $\mathbf{M} := \{M{\subset} \mathbb{R}^3| M {=} \overline{M}\}$, where $\overline{M}$ is the closure of $M$; in the implementation, we represent the models by their boundaries as triangle meshes.
We define a set of CAD modeling operators $\mathbf{O} = \{\mathcal{O}(\theta,\cdot): \mathbf{M} \rightarrow \mathbf{M}\}$, where each applied operator $M' = \mathcal{O}(\theta,M)$ changes the input geometry ${M}$ to a new shape ${M}'$ and is defined by a parameter vector $\theta$ \rev{ that specifies both 
the parts of $M$ to be modified and the corresponding modification parameters}. Note that different operations can have different number and types of parameters. 
In the sketch-based CAD modeling, our primary goal is to interpret a sketch drawn over an existing shape as the corresponding operator with proper parameters that changes the shape to match the 2D sketch.
The overall process is illustrated in Fig.~\ref{fig:test_stage_pipeline}.
Formally, given the current shape ${M}$ and the 2D sketch curves $\mathbf{S}=\{s_i\}$ with known viewpoint $\mathbf{v}\in \mathbb{R}^6$, 
we strive for the mapping $\Phi({M}, \mathbf{S}, \mathbf{v}) = \mathcal{O}(\theta,\cdot)$ such that the image of $\mathcal{O}(\theta,M)$ closely matches $\mathbf{S}$ when viewed according to $\mathbf{v}$.
We use the orthogonal 3D to 2D projection in our approach.
In the following discussion, whenever possible, we omit the parameters of an operator for brevity.

Due to the diversity and infinite variation of operators,
neither the brute-force exhaustive enumeration of all operators and parameters nor the traditional stochastic or energy based optimizations can efficiently solve the inverse problem.
Instead, we approach this problem by using deep learning.
In particular, we train a two-stage neural network that models the mapping $\Phi$, 
where the first stage predicts the operator type and the second stage segments the sketch and context maps into regions, 
on which the specific parameters for the operator are fitted to instantiate the operation.
The key technical challenges are how to design the machine learning models and training tasks such that the inverse mapping is feasible, learned by the neural networks and reliably generalized to real modeling interaction.

We present the definition and parameterization of specific operators in Sec.~\ref{sec:ops}, 
the neural networks and their usage for the inverse mapping in Sec.~\ref{sec:networks}, and how to train the networks for reliable generalization in Sec.~\ref{sec:network_training}.

\section{The Operators}
\label{sec:ops}

\begin{figure}
  \centering
  \begin{minipage}{0.65\linewidth}
  \small{
  \begin{tabularx}{\textwidth}{lX} 
      \toprule
      \multicolumn{2}{c}{\texttt{Extrude}} \\
      \hline
      param: & base face $f$, offset $d$ \\ 
      action: & move $f$ along its normal direction for $d$ \\
      sketch: & edges of the moved $f$ and the extended side edges \\
      \bottomrule
  \end{tabularx}
  }
  \end{minipage}
  \begin{minipage}{0.3\linewidth}
  \centering
  \begin{overpic}[width=0.5\linewidth]{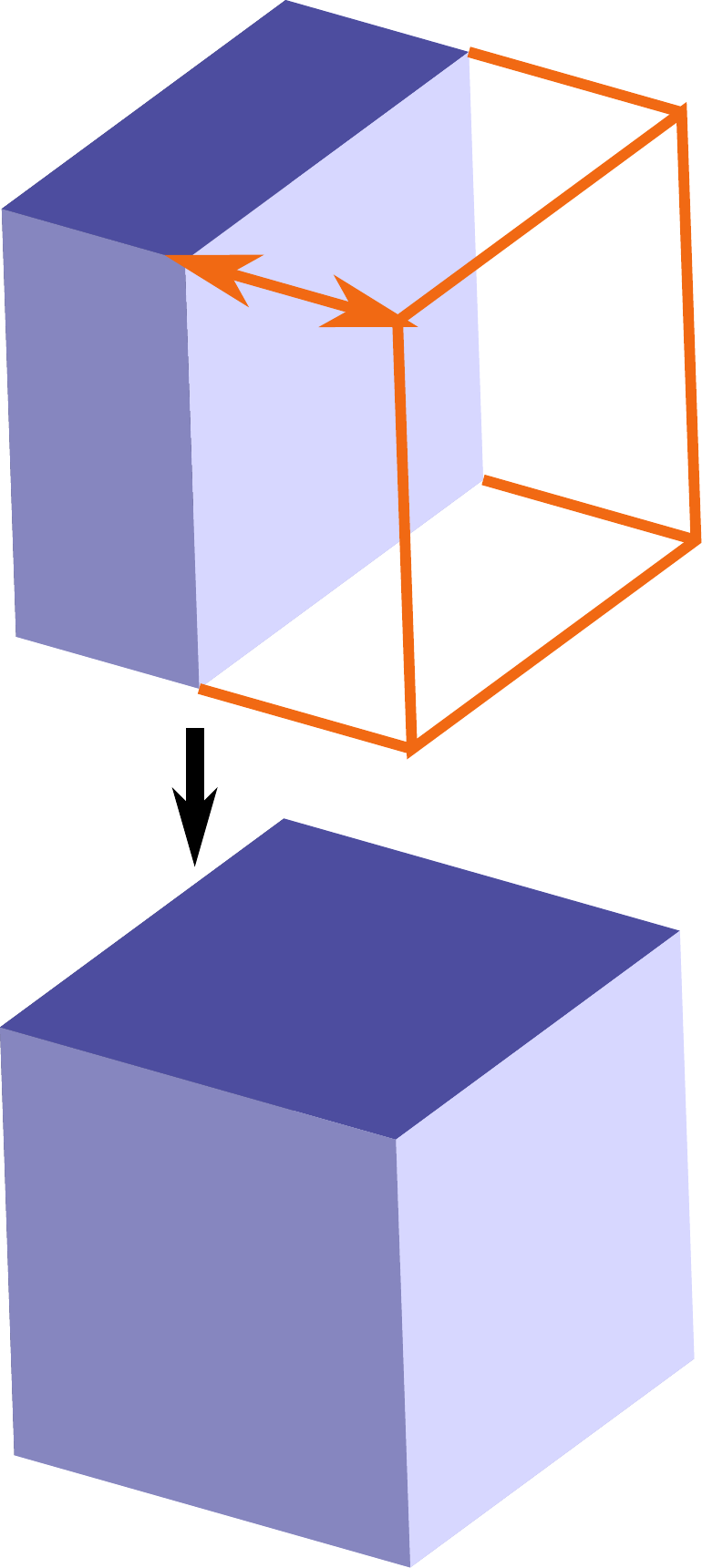}
    \put(16,63){\small $f$}
    \put(19,83){\small $d$}
  \end{overpic}
  \end{minipage}

\begin{minipage}{0.65\linewidth}
	\small{
		\begin{tabularx}{\textwidth}{lX} 
			\toprule
			\multicolumn{2}{c}{\texttt{Bevel}} \\
			\hline
			param: & base face $f$, corner $c$ with an opposite corner $c'$, profile curve $l$ on $f$ \\
			action: & turn $c$ and $c'$ into rounded corners specified by $l$ \\
			sketch: & $l$ and its offset by vector $cc'$ \\
			\bottomrule
		\end{tabularx}
	}
\end{minipage}
\begin{minipage}{0.3\linewidth}
	\centering
	\begin{overpic}[width=\linewidth]{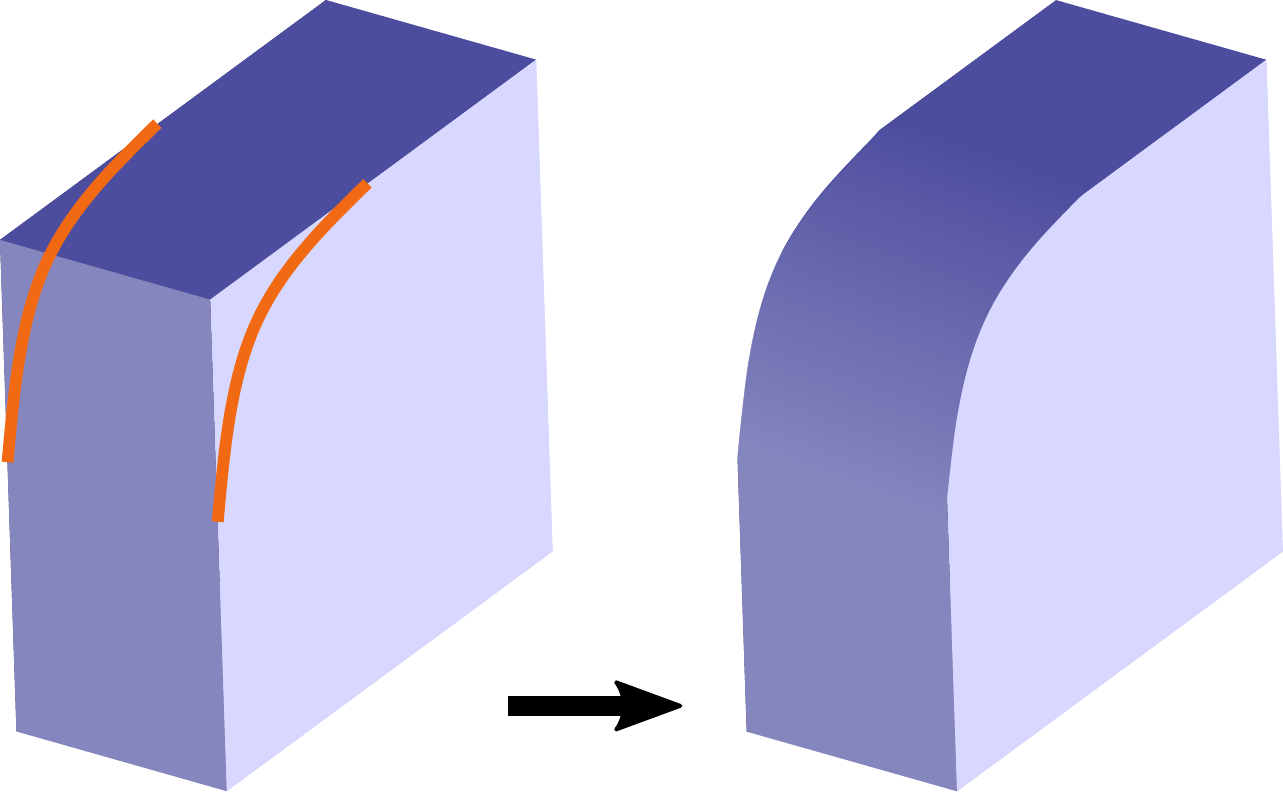}
		\put(30,20){\small $f$}
		\put(29,43){\small $l$}
		\put(12,38){\small $c$}
		\put(-5,42){\small $c'$}
	\end{overpic}
\end{minipage}
\vspace{3mm}

\begin{minipage}{0.65\linewidth}
	\small{
		\begin{tabularx}{\textwidth}{lX} 
			\toprule
			\multicolumn{2}{c}{\texttt{Add/Subtract}} \\
			\hline
			param: & base face $f$, prism base curve $c$, profile length $d$, add/subtract option $o=\pm$ \\
			action: & build a prism with base $c$ and profile edge of length $d$ in the normal direction of $f$, then find the union ($o=+$)/difference ($o=-$) of base shape and the prism \\
			sketch: & edges of the prism \\
			\bottomrule
		\end{tabularx}
	}
\end{minipage}
\begin{minipage}{0.3\linewidth}
	\centering
	\begin{overpic}[width=0.6\linewidth]{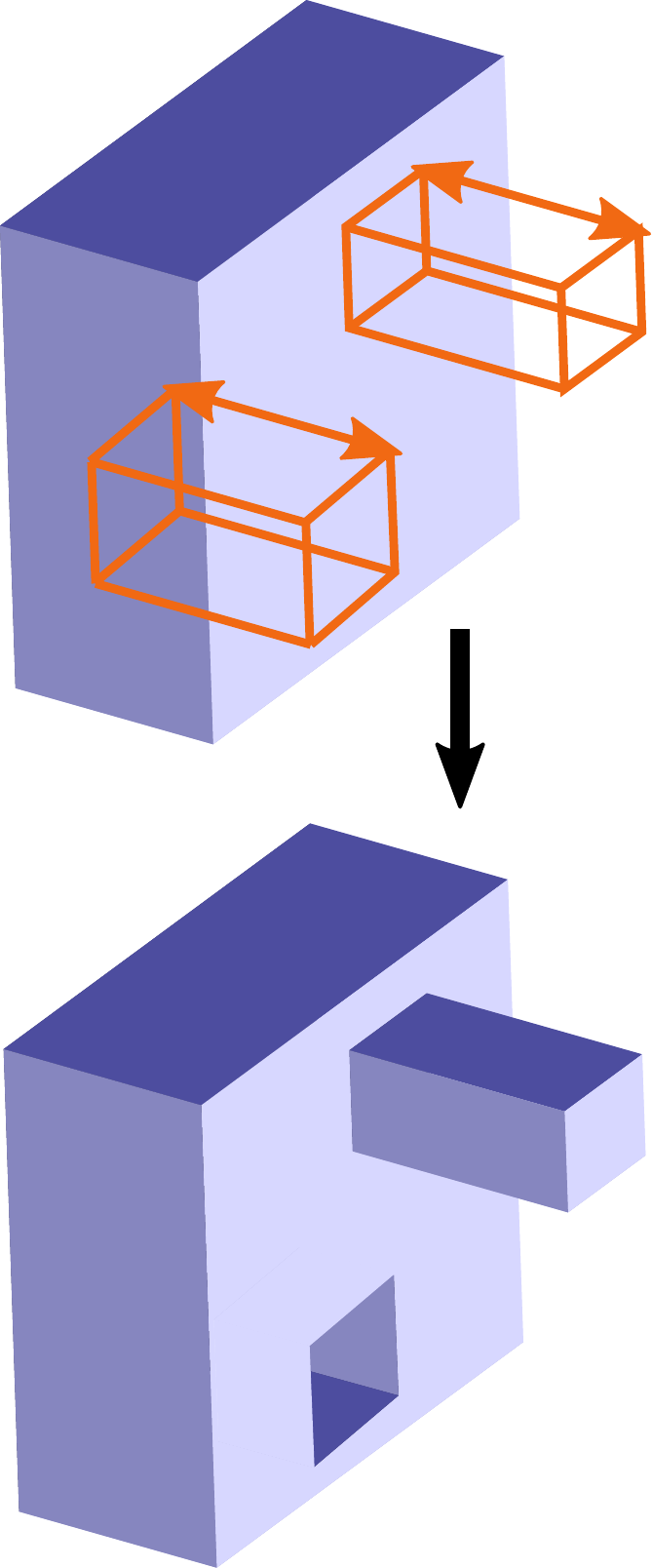}
		\put(28,70){\small $f$}
		\put(32,89){\small $d$}
		\put(18,82){\small $c$}
		\put(44,82){\small $o{=}+$}
		\put(15,75){\small $d$}
		\put(15,55){\small $c$}
		\put(-3,56){\small $o{=}-$}
	\end{overpic}
\end{minipage}
\vspace{3mm}

\begin{minipage}{0.65\linewidth}
	\small{
		\begin{tabularx}{\textwidth}{lX} 
			\toprule
			\multicolumn{2}{c}{\texttt{Sweep}} \\
			\hline
			param: & base face $f$, base/offset circles $c_0,c_1$, profile curve $c_p$, add/subtract option $o=\pm$ \\
			action: & build a swept shape by rolling the profile curve $c_p$ along $c_0,c_1$, then find the union ($o=+$)/difference ($o=-$) of base shape and the primitive \\
			sketch: & circles, profiles of swept shape \\
			\bottomrule
		\end{tabularx}
	}
\end{minipage}
\begin{minipage}{0.3\linewidth}
	\centering
	\begin{overpic}[width=0.8\linewidth]{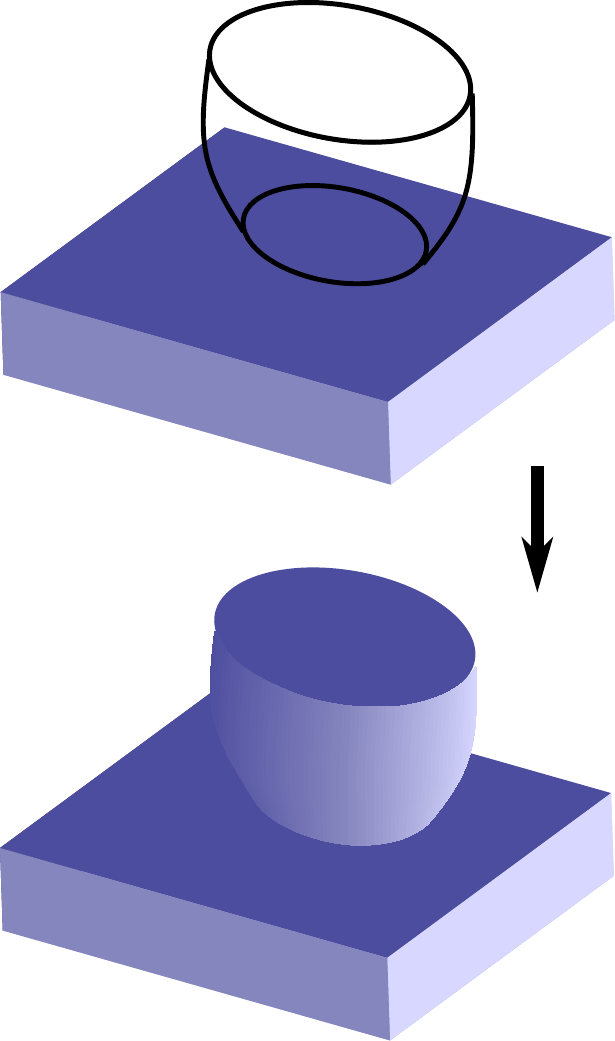}
		\put(30,63){\small $f$}
		\put(15,75){\small $c_0$}
		\put(47,85){\small $c_p$}
		\put(12,95){\small $c_1$}
	\end{overpic}
\end{minipage}
  \caption{{\bf Operators supported in \name.} In each inset, the parameters defining the operator are annotated and the corresponding sketches are shown over the existing shape, while the result of applying the respective operation is shown as the updated shape. }
  \label{fig:operators_illustration}
  \vspace{-4mm}
\end{figure}

In the current system we support the following four operations, i.e. face extrusion, beveling a corner, addition/subtraction of a right polyhedron, and sweeping a cylindrical shape (see Fig.~\ref{fig:operators_illustration} for illustrations).
We choose the four operators because they are widely used both in sketching workflows and CAD modeling, and can already be interleaved to generate complex shapes; 
nonetheless, our system can be easily extended to incorporate more operators as needed.
To fully describe an operator $\mathcal{O}(\theta,\cdot)$, we define its parameters $\theta$, its applied action $M' = \mathcal{O}(\theta,M)$ for given $M$, and the corresponding sketches $\mathbf{S}$ that a user draws to specify it.

\paragraph{\texttt{Extrude}}
Extrusion is the simple offset of a planar face of the 3D shape along the face normal direction.
As shown in Fig.~\ref{fig:operators_illustration}, the parameters defining the operator are the face $f$ to be offset and the distance $d$ along face normal vector for the extrusion, with $d>0$ for pulling out and $d<0$ for pushing in.
The corresponding sketches are the lines extruded and the boundaries of the offset face.

\paragraph{\texttt{Bevel}}
Bevel, also known as \emph{rounding} \cite{eissen2011sketching} in sketching or \emph{fillet} in CAD modeling, is to turn a sharp crease of an object into a smooth and rounded connection.
As shown in Fig.~\ref{fig:operators_illustration}, the operator is defined on the crease connecting two corners $c,c'$, with $c$ residing on the base face $f$; 
the sharp edge $cc'$ is then turned into a smooth connection with profile curve $l$ that rounds $c$ on $f$.
The sketches corresponding to such an operator are the profile curve $l$ on $f$ and its parallel obtained by translation by $cc'$.

\paragraph{\texttt{Add/Subtract}}
The addition or subtraction operator is to place a primitive shape (a prism) over a base shape and compute the union or difference of the two shapes.
The parameters are used to designate the base face $f$ to place the primitive, and to define the primitive shape by specifying its base curve $c$ as one of triangle, quadrilateral, pentagon, or hexagon, as well as its profile curve that is always parallel to the base face normal with length $d$.
In addition, the option $o$ of union (for addition) or difference (for subtraction) between the base shape and the primitive is specified.
The corresponding sketches are simply depicting the primitive shape by highlighting its feature curves.

\paragraph{\texttt{Sweep}.}
Sweeping a curved profile line along two circular rails is another commonly used operation in CAD modeling, which also appears frequently in industrial design sketching in the form of horizontal ellipses joined by a vertical section (see Fig.~\ref{fig:operators_illustration}).
Similar to \texttt{add/subtract}, the swept shape is combined with the base shape by either union or subtraction; 
therefore, the \texttt{sweep} operation can be seen as a special \texttt{add/subtract} where the primitive shape is a swept cylindrical shape.
The parameters to define the \texttt{sweep} operation consist of the base and offset circles, and the profile curve whose two ends lie on the two circles.
There is also the union and difference option to specify the combination with base shape.
The corresponding sketches simply show the swept shape through its two circular ends and a pair of profile curves.
The \texttt{add/subtract} and \texttt{sweep} operators are denoted as \texttt{Add/SubtractPolyhedron} and \texttt{Add/SubtractSweepShape} respectively in operation sequences (see Figs.~\ref{fig:teaser},~\ref{fig:test_stage_pipeline} and~\ref{fig:train_stage_pipeline}) for distinction.

\paragraph{Extension to more operators.}
One can follow the above examples to define new operators.
In general, the parameters of the operator should be minimal but complete in defining its actions without ambiguity.
The corresponding sketches should be concise and capture the important features of the operation.
All these designs will impact the machine learning models used for recovering the operator instance from sketches, as discussed later in Sec.~\ref{sec:networks}.

\paragraph{Protocols for CAD modeling.}
A protocol file is a serialization of the modeling steps.
It consists of the full set of parameters specifying the operators that are applied in sequence to obtain the final shape.
A protocol can be saved, loaded, edited, and reused for more complex modeling tasks.
Illustrations of a protocol as the sequence of operations it contains are shown in Figs.~\ref{fig:teaser},~\ref{fig:test_stage_pipeline},~\ref{fig:train_stage_pipeline}.
More protocol texts for generating models shown in Fig.~\ref{fig:result_gallery} can be found in the supplemental material.

\paragraph{Implementation of operator actions.}
In our current implementation, we represent the 3D solid models by their boundaries as triangle meshes, but always maintain a set of planar polygonal faces that are made of adjacent triangles of coplanarity, by flooding across mesh edges with tight dihedral angle thresholding ($<1^\circ$).
When applying any of the operators defined above which require planar bases, the base face is selected as one of the planar polygons,
and the action is carried out by computing the appropriate Boolean operation between the sketched primitive and the base mesh, using CGAL~\cite{cgal}.
While for \texttt{extrude,add/subtract,sweep} the sketched primitives are clear, for \texttt{bevel}, we construct a prism whose base face is defined by connecting $c$ and $l$ and whose profile edge is $cc'$ (Fig.~\ref{fig:operators_illustration}), and subtract it from the base shape.
In the future, when extending to operators applied on curved faces, we consider upgrading our underlying geometry representation to more flexible ones, e.g. NURBS (Sec.~\ref{sec:future_work}).

\section{Context Driven Sketch Interpretation}
\label{sec:networks}

After sketching an operation in the modeling session, 
there are three steps taken to interpret the current sketch $\mathbf{S}$: 
the recognition of the operator type $\mathcal{O}$ by an classification network,
the extraction of the individual regions from the input maps by the segmentation network for the operator type $\mathcal{O}$,
and the recovery of parameters $\theta$ defining the specific instance $\mathcal{O}(\theta, \cdot)$ by counting and curve fitting.
Finally, the regressed operator is applied to the existing geometry to carry out the modeling intention of the user.

For all networks, the input is the concatenation of three maps, all of spatial size $256\times256$:
the sketch map $S$, with $S(x,y)=1$ for the stroke pixels $(x,y)$ and $S(x,y)=0$ otherwise,
and the local context maps $D,N$ encoding depth and normal, obtained by rendering the existing geometry along the sketched viewpoint $\mathbf{v}$.
The viewing frustum for generating the maps is twice the size of the sketch bounding box, to ensure the user input is well covered.
For the depth map, $D(x,y)\in[0,1]$ is the normalized depth value for a foreground pixel $(x,y)$ and $D(x,y)=0$ to indicate background pixels.
We normalize the depth map linearly such that the farthest depth is mapped to 0 and the closest depth to 1,
thus removing the dynamic range of possible depth values to ease learning.
For the normal map, $N(x,y)\in \mathbb{R}^3$ is the normal vector that is first transformed into the 3D camera space and then shifted by $(1,1,1)$ for a foreground pixel, and $N(x,y)=(0,0,0)$ for background pixels.

\subsection{Operator classification}

The classification network is a CNN with alternative layers of convolution and pooling that finally outputs the probabilities for the operator types that the input sketch represents; see supplemental material for the detailed structure.
The training loss is the weighted cross entropy:
\begin{equation}
    \mathcal{L}_{cls}(S,D,N) = - w_\mathcal{O}\log{(P_\mathcal{O})},
    \label{eq:classification_loss}
\end{equation}
where $\mathcal{O}$ is the ground truth operator type for the input training sample, 
the class weights $\left(w_\mathcal{O'}\right)_{\mathcal{O'}\in \mathbf{O}}$ are computed by normalizing the inverse type frequency vector $\left(\frac{1}{N_\mathcal{O'}}\right)_{\mathcal{O'}\in \mathbf{O}}$, with $N_\mathcal{O'}$ the number of training samples of type $\mathcal{O'}$, 
and $P_\mathcal{O}$ is the predicted probability of the operator being of type $\mathcal{O}$.
We use weights for different operation types to avoid potential statistical bias caused by their contrastive frequencies in the training set, as discussed in Sec.~\ref{sec:network_training}.

\subsection{Operator regression}
Rather than directly regressing the parameters,
we solve the regression in two steps:
first, we use deep neural networks to segment the sketch and context maps into regions corresponding to the defining structures of the operators, and second, we fit operation parameters to the detected regions using counting, searching and optimization procedures.
The benefits of such a two-step regression, are that the  networks are only required to learn the single-modality segmentation tasks, 
which is considerably more tractable than brute-force regression of diverse operator parameters, 
and that the parameter fitting is robust to inaccuracies of network predictions. Further, this design choice facilitates  generalization across different operations. 
In contrast, by trying to regress directly the various parameters of an operation, we face several difficulties: to recover the extrusion and offset distances from 2D images has the inherent scale ambiguity, the number of base polygon sides of the \texttt{add/subtract} operation is changing and needs complex network structures to accommodate, and the regression of curved strokes requires fixed Bezier or spline parameterization, while in our case we can choose suitable representations to do the curve fitting.

\paragraph{The general network structure.}

Each of the segmentation networks is a U-Net that outputs two maps through two decoder branches: the probability map $F$ of base face, and the curve segmentation map $C$, both of spatial size $256\times 256$ and channel width 1;
details of network structures are provided in the supplemental material.
To train the network, the loss function is in the following general form:
\begin{equation}
    \mathcal{L}_{reg}(S,D,N) = \frac{1}{256^2}\|F - \widetilde{F}\|^2 + \frac{1}{|\widetilde{M}|}\|\widetilde{M}\odot (C - \widetilde{C})\|^2 ,
    \label{eq:general_regression_loss}
\end{equation}
where maps with $\sim$ are ground truth or precomputed maps, i.e., $\widetilde{F}$ is the ground truth base face map with $\widetilde{F}(x,y)=1$ for foreground pixels and zero otherwise, 
$\widetilde{C}$ is the ground truth stroke map,
and $\widetilde{M}$ is the corresponding stroke pixel mask.
$\odot$ is the component-wise product, and $|\cdot|$ sums the map pixel values.

Given the predicted face map, we find the base face $f$ by counting. 
To be specific, we first binarize the face map $F$ by threshold 0.5, then render the face ID map of existing geometry $Id(x,y)\in\{f_i\}$, and finally find the face $f = f^*$ with the highest accumulated probability, computed as $f^* = \argmax_{f_i}{\sum_{Id(x,y)=f_i}{F(x,y)}}$.

Different operators have their specific curve segmentation maps $C$ and $\widetilde{C}$.
The principle for designing the curve maps is that \textit{the input-output pair should be a learnable mapping without strong ambiguity}.
Next, we present the details of regression for each operator.

\paragraph{Extrude regression.}
We specify the ground truth extrusion curve segmentation map in this way: $\widetilde{C}(x,y) = 1$ for pixels of the offset curve, and $\widetilde{C}(x,y) = 0$ for profile curve pixels (see Fig.~\ref{fig:operators_illustration}).
Correspondingly, given the network predicted curve map $C$, 
we find the map of offset curve as $C_o(x,y) := (C(x,y) > 0.5)\wedge (\widetilde{M}(x,y)=1)$, and the map of profile curves as $(C(x,y) \leq 0.5) \wedge (\widetilde{M}(x,y)=1)$. 

Having classified the pixels, we find the extrusion distance $d$ by line search.
In particular, the edges of the base face, denoted as $\partial f$, are extruded along normal direction $\mathbf{n}_f$ for $d$ to match $C_o$.
The linear search has a fine step size $\sigma=0.0075$ and search range $[-1.5,1.5]$, whereas the initial shapes have unit diagonal bounding box length.
Note that by including the negative search range, we allow pushing the base face inside the model as well.
We define the matching distance of the extruded face edges and the offset curve map, as $dist(d) := \sum_{\mathbf{p}\in\partial f}{\min_{C_o(\mathbf{q})=1}\|\pi_{\mathbf{v}}(\mathbf{p}+d \mathbf{n}_f) - \mathbf{q}\|}$, where $\mathbf{p}$ samples face edges uniformly by arc length,  $\mathbf{q}$ ranges over image pixels, and $\pi_{\mathbf{v}}: \mathbb{R}^3\rightarrow\mathbb{R}^2$ is the projection function of the current view.
The line searched $d$ with minimum $dist(d)$ is the regressed extrusion distance.

\paragraph{Bevel regression.}

The ground truth curve segmentation map for the \texttt{bevel} network encodes the two curves $l$ and $l'$ (see Fig.~\ref{fig:operators_illustration}) in this way: 
$\widetilde{C}(x,y)=1$ for pixels of $l$ and $\widetilde{C}(x,y)=0$ for pixels of $l'$.
Correspondingly, we find the predicted base face curve map as $C_l(x,y) := (C(x,y) > 0.5)\wedge (\widetilde{M}(x,y)=1)$, and the map of $l'$ as $(C(x,y) \leq 0.5) \wedge(\widetilde{M}(x,y)=1)$.

Assuming the profile curve is drawn in one stroke, we find the stroke corresponding to $l$ by counting.
Let $s^*$ be the stroke with the highest accumulated probability: 
$s^* := \argmax_{s_i\in \mathbf{S}}{\sum_{\mathbf{p}\in s_i}{C_l(\mathbf{p})}}$, 
where $\mathbf{p}$ uniformly samples $s_i$ in the screen space.
We then fit a cubic Bezier curve as $l$ to match $s^*$ as it is back projected onto the plane of $f$.
Given $f$ and $l$, we determine the corner $c$  as the shared vertex of the two edges of $f$ which intersect with $l$.
Once we have $c$, the opposite corner $c'$ is found easily.

\paragraph{Add/subtract regression.}

\begin{figure}
	\centering
	\begin{overpic}[width=0.8\linewidth]{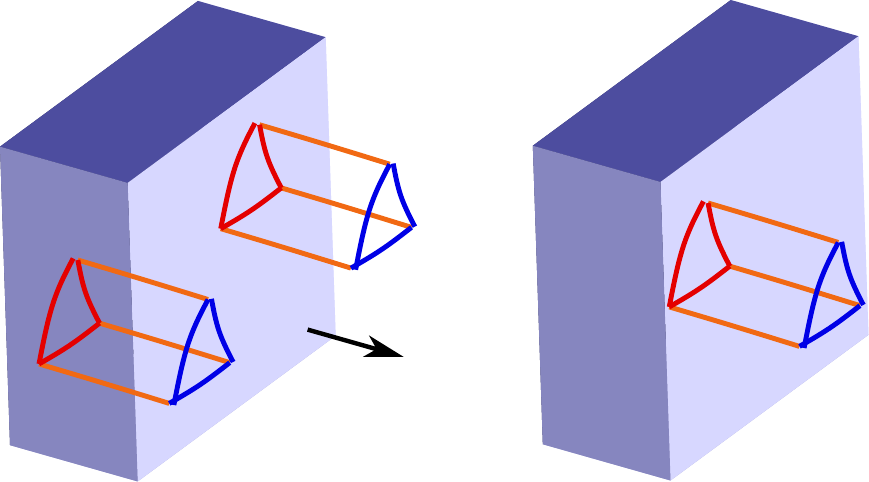}
	    \put(40,40){\small \texttt{add}}
	    \put(3,5){\small \texttt{subtract}}
	    \put(90,32){\small \texttt{add/subtract}}
	    \put(38,11){\small $\mathbf{n}_f$}
		\put(27,3){\small (a)}
		\put(90,3){\small (b)}
	\end{overpic}
	\vspace{-3mm}
	\caption{
	{\bf Handling ambiguity between add versus subtract.} 
	The ambiguity of distinguishing base and offset curves for the \texttt{add/subtract} operator. (a) cases without ambiguity, as only one of the red and blue curves intersects with the base face. 
	(b) ambiguous case that can be \texttt{add} or \texttt{subtract}, with the base curve being either red or blue.
	Instead of segmenting the base and offset curves, we regress two curves along the face normal direction (red first, blue second), thus removing ambiguity.}
	\label{fig:addsub_ambiguity}
	\vspace{-3mm}
\end{figure}

\begin{figure*}
    \centering
    \includegraphics[width=\linewidth]{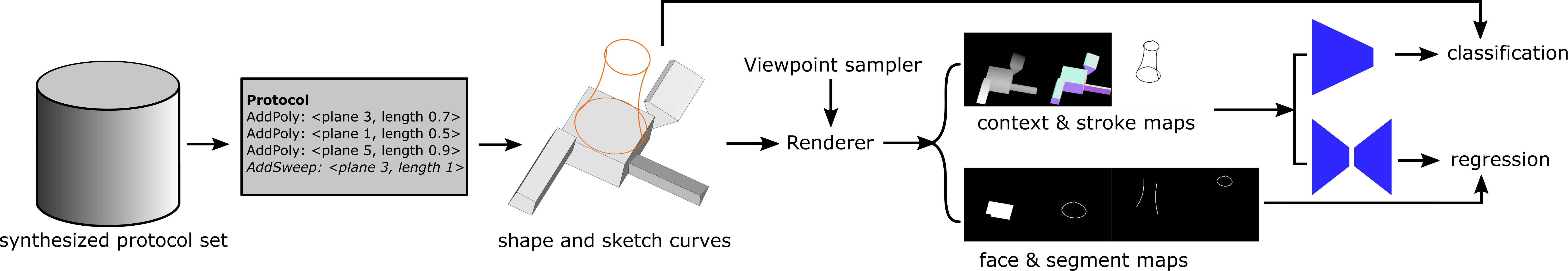}
    \vspace{-7mm}
    \caption{{\bf \name at training time.} We synthetically generated 10k protocols of diverse lengths for procedurally generating 40k training shapes. For each protocol, we execute it up to the last operation, for which the sketch curves are built and overlaid on the built shape. The sketch curves and existing shape are rendered in proper viewpoints to generate the input sketch and local context maps, as well as the ground truth face and curve segmentation maps, which are used to train the operator classifier and the corresponding segmentation network. }
    \label{fig:train_stage_pipeline}
    \vspace{-3mm}
\end{figure*}

For the \texttt{add/subtract} operator, there is ambiguity with outputting the base and offset curves directly, as is illustrated in Fig.~\ref{fig:addsub_ambiguity}.
To remove this ambiguity, we instead regress the two curves as ordered along the face normal direction and named the start and end curves, respectively.
The ground truth curve map is given by 
$\widetilde{C}(x,y)=0$ for the start curve pixels, 
$\widetilde{C}(x,y)=1$ for the profile curve pixels, 
and $\widetilde{C}(x,y)=2$ for the end curve pixels.
Therefore, we obtain the predicted maps for the start curve $C_s(x,y) := (C(x,y)\leq 0.5) \wedge (\widetilde{M}(x,y)=1)$, the profile curve $C_p(x,y) := (0.5 < C(x,y)\leq 1.5)\wedge (\widetilde{M}(x,y)=1)$, and the end curve $C_e(x,y) := (C(x,y)> 1.5) \wedge (\widetilde{M}(x,y)=1)$.

The \texttt{add/subtract} operation has more complex parameters than \texttt{extrude} or \texttt{bevel} (see  Fig.~\ref{fig:operators_illustration}), the recovery of which also involves more steps:
we first classify the strokes according to the predicted curve map, then fit 2D curves to the strokes and determine the add/subtract option, and finally back project the base curve to the 3D base face and recover the prism length by line search.

Again, we classify strokes by pixel counting.
For a stroke $s_i$, let its likelihood of being starting curve as $L_s(s_i) := \sum_{\mathbf{p}\in s_i}{C_s(\mathbf{p})}$, where $\mathbf{p}$ samples $s_i$ uniformly, and similarly we have $L_p(s_i), L_e(s_i)$ for the likelihood of being profile and end curves, respectively; 
the curve type of $s_i$ is the one with largest likelihood.

We assume each of the profile curves is drawn by one stroke; therefore the number of profile strokes tells the $N$-gon of the prism base.
We then find the end points of the profile curves grouped into the beginning set and the end set, which are used as the initial guess for fitting $N$-gons to the starting and ending strokes through iterative closet point method, respectively.

To determine the add/subtract option $o$, we check the intersections of fitted polgyons with the base face $f$.
If the start polygon $P_s$ intersects $f$, we have $o=+$ the addition.
Otherwise if the end polygon $P_e$ intersects $f$, we have $o=-$ the subtraction.
However, if none of the two intersects with $f$, the sketch is regarded erroneous with no matching operator instance.
The user is alerted with this failure.
For the ambiguous case shown in Fig.~\ref{fig:addsub_ambiguity}(b), the above procedure implies the default addition option, and the user can switch it manually if needed (Sec.~\ref{sec:system}).

Finally, the 2D base polygon, i.e., $P_s$ for addition and $P_e$ for subtraction, is back projected onto the plane of $f$ to obtain the 3D prism base polygon, and the prism length $d$ is obtained by line searching the 3D base polygon along normal direction to match the pixels of the offset end.
The line search process replicates that for extrusion.

\paragraph{Sweep regression.}

Since the \texttt{sweep} operator is a special case of the \texttt{add/subtract} operator, the curve maps are the same for both operators.
The fitting of parameters is much like \texttt{add/subtract} operator as well, with minor differences in curve fitting. Note that we restrict the \texttt{sweep} operations to circular cross sections. 

The start and end curves are fitted as ellipses to the corresponding curve maps. 
After having determined the base curve and add/subtract option, 
the ellipse is back projected to the base face plane as the base circle $c_0$ (see  Fig.~\ref{fig:operators_illustration}).
The offset $d$, defined as the distance between the two circle centers, is again found by line search as done in extrusion, except to match the base circle center to the offset curve center in this case.
The offset circle $c_1$ is then obtained by back projecting the offset ellipse to the translated base face plane by distance $d$.

To recover the profile curve, we first determine the 3D plane it lies on. 
For one of the 2D profile strokes, it has an intersection point with each of the two ellipses. 
The intersection points are lifted to 3D following the ellipse-circle back projection.
The centers of $c_0,c_1$ and any of the two intersection points together determine the 3D plane that the profile curve resides in.
We fit a cubic Bezier curve inside the plane to the profile stroke points, to obtain the profile curve.
Finally, if we detect that the profile curve to be nearly linear and that the two circles have similar radii, we rectify the swept shape to be a cylinder.

\begin{figure}[b!]
    \centering
    \includegraphics[width=0.7\linewidth]{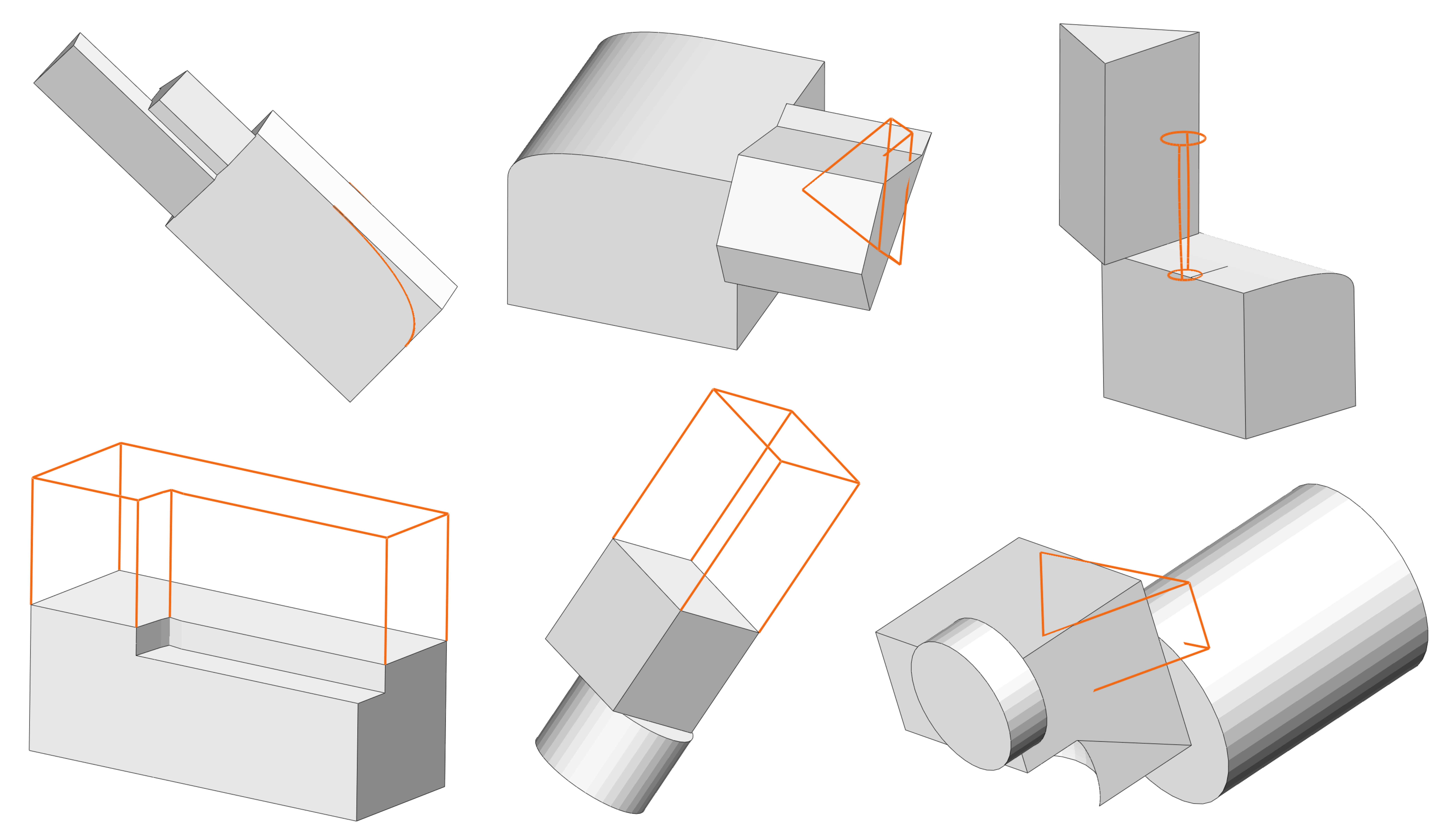}
    \vspace{-3mm}
    \caption{{\bf Procedurally generated training set.} Sample synthesized shapes and next step sketch by randomized combination of operations. Within only four steps, very complex shapes can already be created.}
    \label{fig:sample_synthetic_shapes} 
\end{figure}

\section{Training data generation}
\label{sec:network_training}

To train the networks for robust performance on real sketching interactions, 
we need a large-scale data set that covers the possible variations.
Thanks to the procedural nature of CAD modeling, 
we can generate the training data by synthesizing diverse procedures (Fig.~\ref{fig:train_stage_pipeline}).
The training data generation therefore consists of two steps, the modeling sequence generation and the sketch image rendering, as discussed below.

\paragraph{Sequence generation}

Given a set of CAD modeling operations $\{\mathcal{O}_i\}$, we generate training data that allows the network to learn to infer from 2D sketches the corresponding operations robustly, while avoiding the prohibitive enumeration of the infinite space of all possible 3D models and configurations of operations.
Our key observation for achieving this goal is that while in theory one part of a 3D model can potentially be connected with every other part of the model, 
it is the local context that influences the part geometry the most and therefore provides the dominant cue for interpreting its 2D sketch properly.
Based on this observation, we only need to extensively enumerate the local combinations of different operations producing diverse model variations to train the network.
Thus in practice, for each operation $\mathcal{O}$, in addition to its own parametric variations, we search for a sequence $\{\mathcal{O}_{1},\cdots,\mathcal{O}_{m}\}$ of random operations, that are applied before the operation, i.e.,  $\mathcal{O}\circ\mathcal{O}_{m}\circ\cdots\circ\mathcal{O}_{1}$, to simulate the local context variations.
Indeed, we find that with $0\leq m \leq 3$, there can be very complex combinations and shapes generated; some examples are shown in Fig.~\ref{fig:sample_synthetic_shapes}.

To balance complexity, for each sequence length $m+1\in[1,2,3,4]$, we generate 10k protocols, thus 40k shapes in total.
For sequences of each length, the last operator $O$ has a fixed frequency for different types, i.e., $1:1:4:2$ for \texttt{extrude}, \texttt{bevel}, \texttt{add/subtract} and \texttt{sweep}; the ratios are chosen to account for the different complexities of the four operators, allocating more samples for \texttt{add/subtract} and \texttt{sweep} which have more degrees of freedom. 
In addition, we generate 10k protocols with the same distribution for testing the networks.
Note that since we weight the different operation types by their inverse frequencies in the dataset (Eq.~\ref{eq:classification_loss}) for training the classification network, such a non-uniform distribution does not cause bias for operation recognition.

While always starting with a base box shape, we randomize each operation instance in a synthesized sequence to cover sufficient geometric variations while avoiding degeneracy. 
This includes, for example, selecting a random planar face from the existing shape as the base face, applying offsets sampled in a large range, generating base polygons or circles that have centers positioned randomly inside the base face region, and polygons and profile curves that are perturbed without self-intersection.

\paragraph{Sketch rendering}
We design the rendering process to mimic how real sketch drawing looks like.
To render the corresponding sketch and context maps of an operator in the generated sequence, 
we randomly sample informative views around the base face of the operator, with view directions forming angles in the range of $[20^\circ,80^\circ]$ with the face normal.
The viewing frustum is centered around the sketch curves, and further scaled by a random factor in range $[1.6,2.4]$ to create different zooming effects.
We filter out the views where for the extrusion operation, the offset curve is occluded for more than $20\%$, or for the other operations, the base curve is occluded for more than $20\%$, as such viewpoints are unnatural to take in real sketching.
The 3D curves of an applied operator instance are first projected onto the 2D camera space, then perturbed at the endpoints randomly by a Gaussian noise, and finally smoothed a little for regularization, which reproduces the style of rough freeform sketching.


\begin{figure}[b!]
    \centering
    \includegraphics[width=0.7\linewidth]{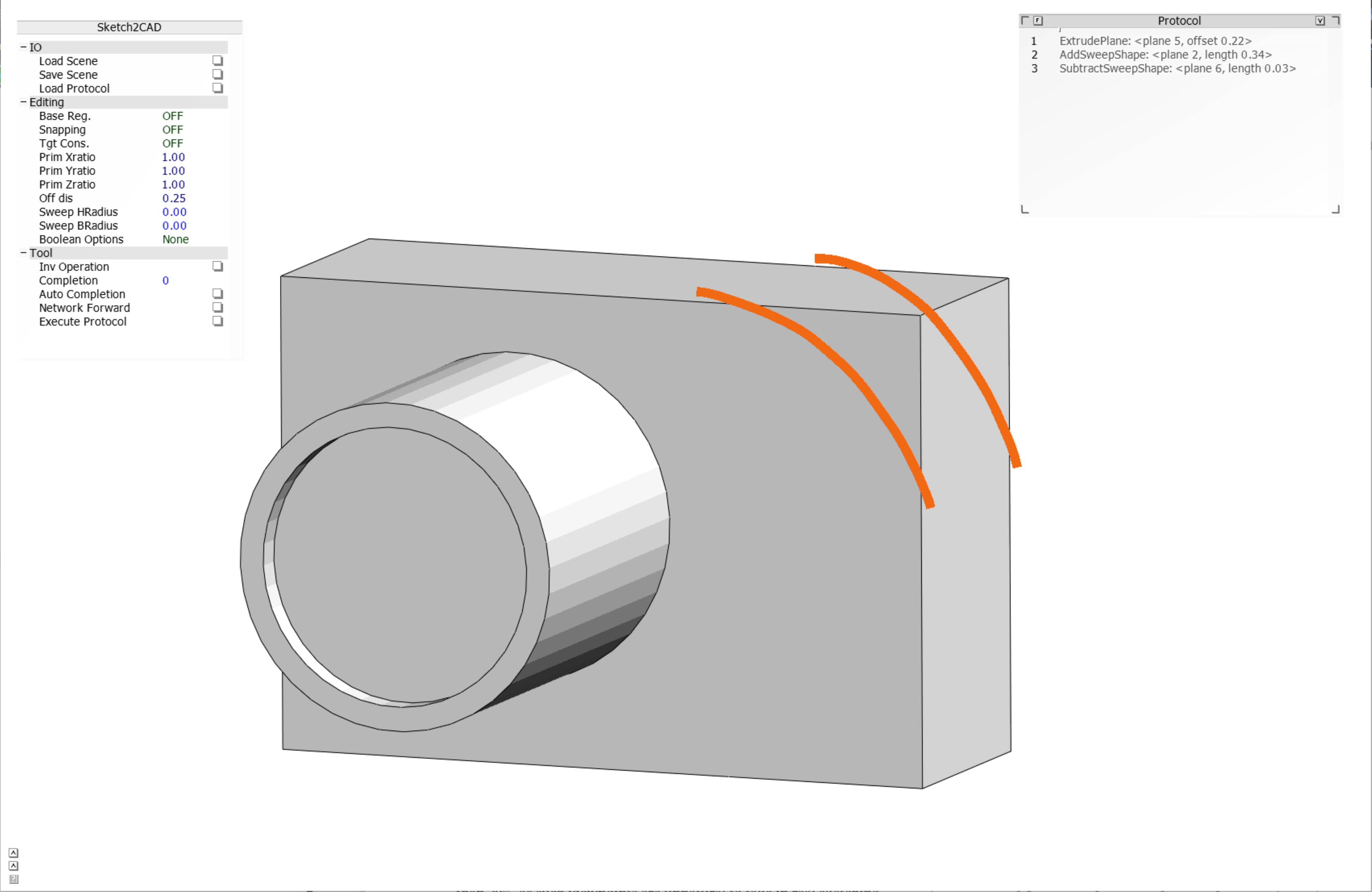} 
    \vspace{-3mm}
    \caption{{\bf \name UI.} A screenshot of our prototype implementation. The tool features freeform and interactive sketching over a 3D shape in the central canvas, the editing of recovered operation parameters on the left, and the illustration of the operation sequence on the right.}
    \label{fig:screenshot} 
\end{figure}

\section{The Modeling System}
\label{sec:system}

We build a prototype modeling tool to demonstrate our approach.
The tool features interactive modeling with an user interface that allows sketching in 2D and instant feedback viewed freely in 3D. 
A screenshot of the our tool is shown in Fig.~\ref{fig:screenshot}.
Please refer to the supplemental video for real time sketching and modeling sessions.

Besides sketching, the tool allows the user to save, load, replay and edit the sequence of operations stored in protocol files, thus fully demonstrating the power of the procedural CAD modeling paradigm.
To assist the easy sketching of \emph{precise} CAD models, our system also implements techniques like the
regularization of sketched curves, the tuning of operation parameters, and auto-completion by replicating sketched primitives through symmetry, as detailed next.

\begin{figure}[t!]
    \centering
    \begin{overpic}[width=0.9\linewidth]{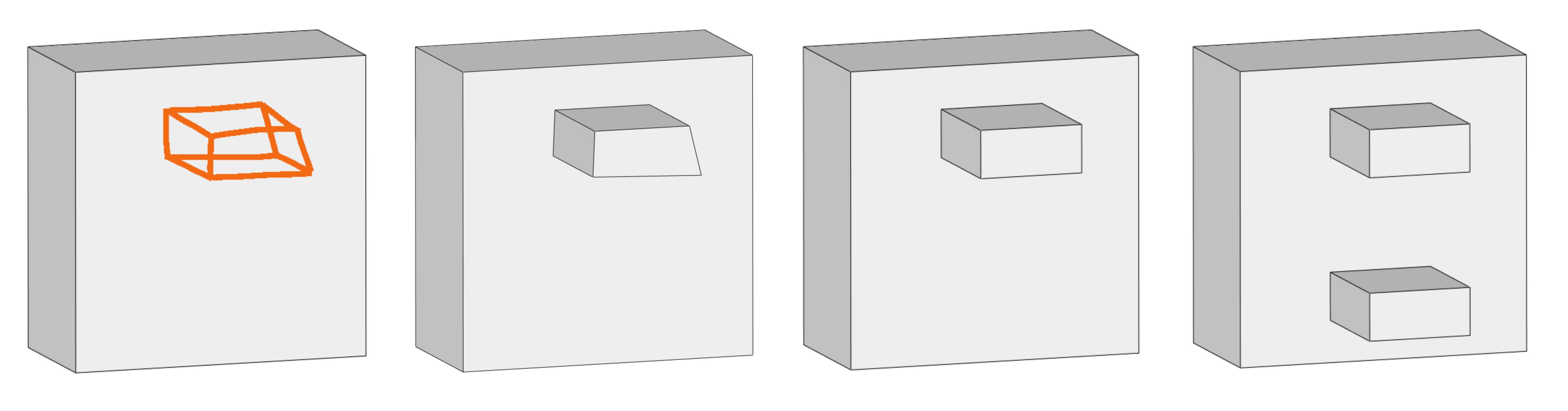}
	\put(13,0){\small (a)}
	\put(38,0){\small (b)}
	\put(63,0){\small (c)}
	\put(88,0){\small (d)}
	\end{overpic}
    \vspace{-3mm}
    \caption{{\bf Stroke regularization and auto-completion in Sketch2CAD. } (a)-(c): the corners of the sketched quadrangle are detected to be close to right angles, and automatically regularized to form a rectangular base polygon. (c)-(d): auto-completion by reflecting the primitive against the horizontal cross section of the base shape.}
    \label{fig:regularization}
    \vspace{-3mm}
\end{figure}

\begin{figure}[b!]
    \centering
    \begin{overpic}[width=0.9\linewidth]{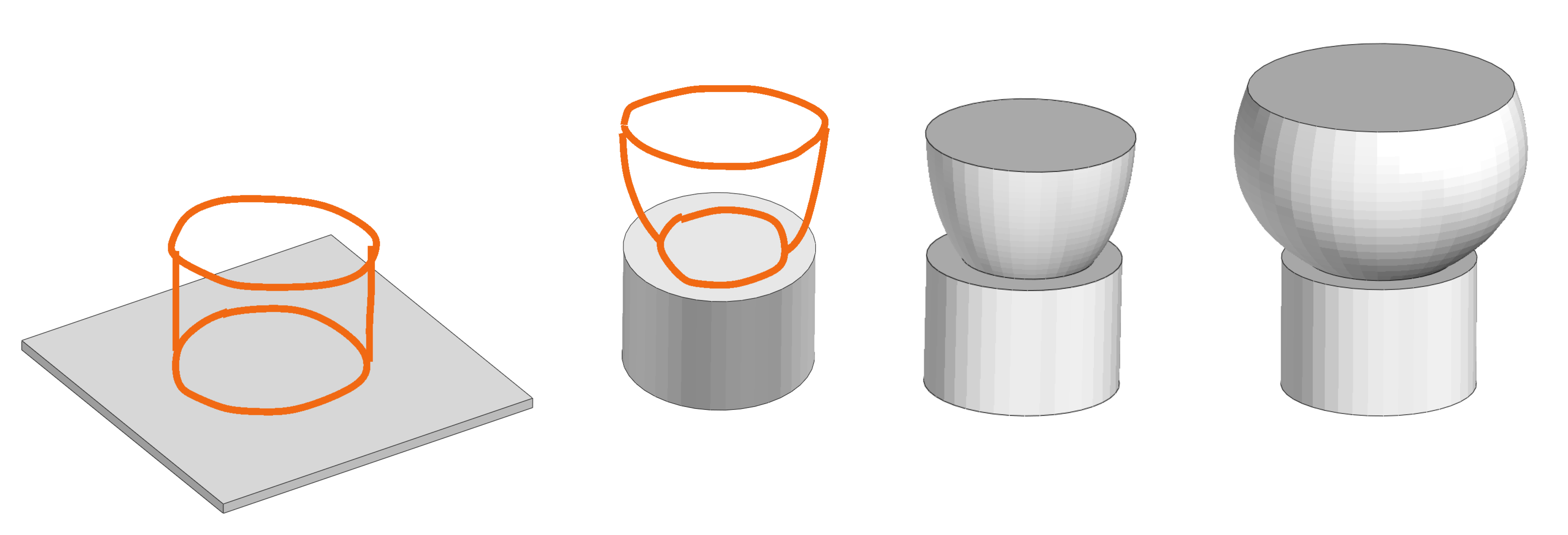}
		\put(3,3){\small (a)}
		\put(44,3){\small (b)}
		\put(64,3){\small (c)}
		\put(87,3){\small (d)}
	\end{overpic}
	\vspace{-4mm}
    \caption{\textbf{Tuning the parameters of operations in \name}. (a)-(b): a cylinder is sketched onto the base box, but only the cylinder is kept. (b)-(c): a swept shape is added. (c)-(d): the offset distance between the two circles of the swept shape, as well as the top circle radius, are enlarged by tuning their parameter values. }
    \label{fig:parameter_tuning} 
    \vspace{-3mm}
\end{figure}

\begin{figure}[t!]
    \centering
    \begin{overpic}[width=\linewidth]{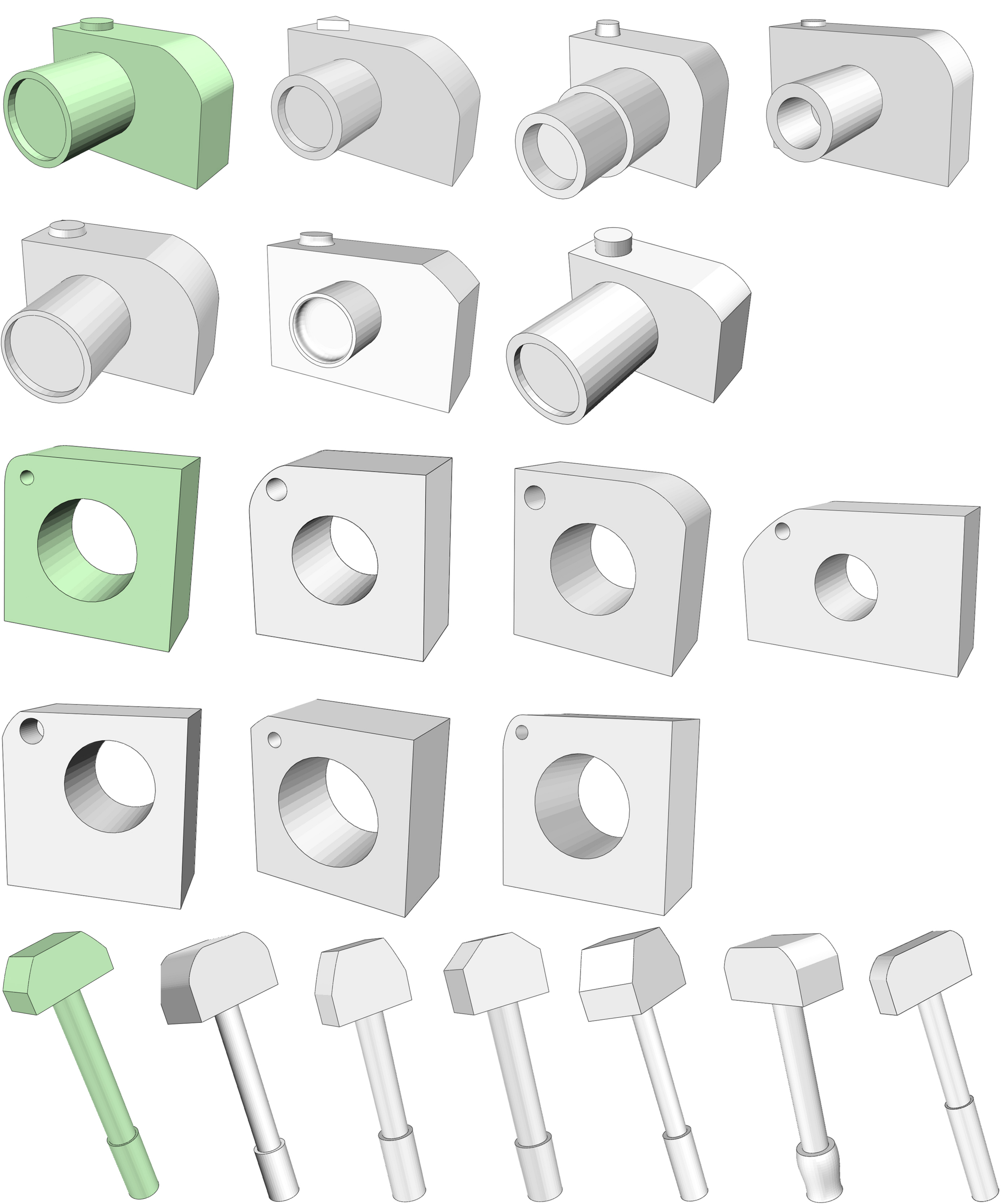}
    \put(3, -2) {\footnotesize{Ref.}}
    \put(15, -2) {\footnotesize{P1, 12mins}}
    \put(28, -2) {\footnotesize{P2, 14mins}}
    \put(40, -2) {\footnotesize{P3, 10mins}}
    \put(52, -2) {\footnotesize{P4, 10mins}}
    \put(64, -2) {\footnotesize{P5, 6mins}}
    \put(76, -2) {\footnotesize{P6, 10mins}}
    \put(13, 63) {\footnotesize{Ref.}}
    \put(33, 63) {\footnotesize{P1, 10mins}}
    \put(53, 63) {\footnotesize{P2, 8mins}}
    \put(72, 63) {\footnotesize{P3, 14mins}}
    \put(13, 42) {\footnotesize{P4, 10mins}}
    \put(33, 42) {\footnotesize{P5, 8mins}}
    \put(53, 42) {\footnotesize{P6, 8mins}}
    \put(13, 99) {\footnotesize{Ref.}}
    \put(33, 99) {\footnotesize{P1, 20mins}}
    \put(53, 99) {\footnotesize{P2, 23mins}}
    \put(72, 99) {\footnotesize{P3, 15mins}}
    \put(13, 81) {\footnotesize{P4, 10mins}}
    \put(33, 81) {\footnotesize{P5, 10mins}}
    \put(53, 81) {\footnotesize{P6, 4mins}}
    \end{overpic}
    \caption{{\bf User gallery.} We asked 6 participants to reproduce 3 reference shapes, shown in green. All participants completed these modeling tasks in 5 to 20 minutes and achieved a close match to the reference.}
    \label{fig:user_gallery}
    \vspace{-5mm}
\end{figure}

\begin{figure*}[t!]
    \centering
    \begin{overpic}[width=\linewidth]{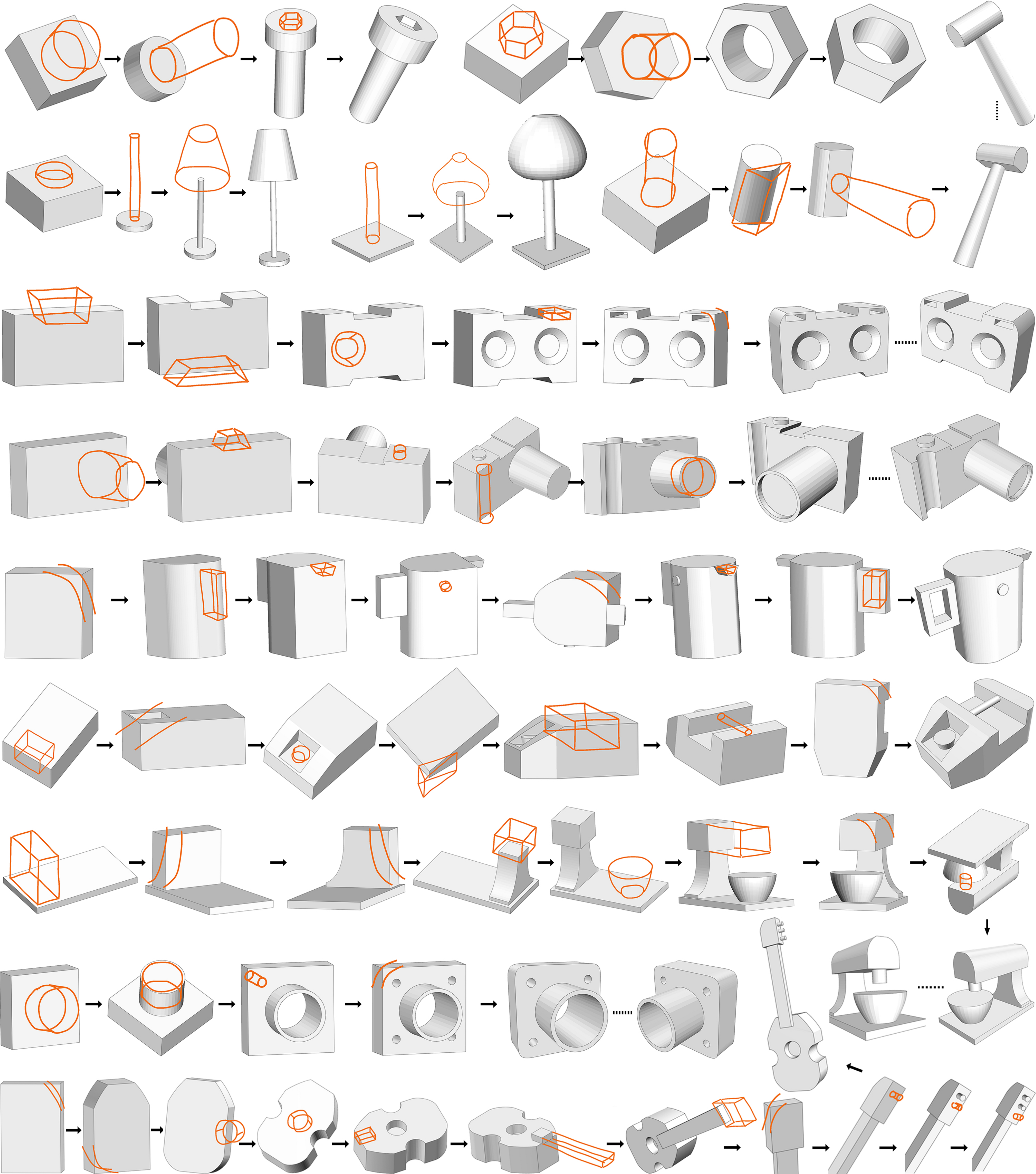}
    \put(1, 98) {\small a}
    \put(1, 86) {\small c}
    \put(39, 98) {\small b}
    \put(30, 86) {\small d}
    \put(52, 86) {\small e}
    \put(1, 74) {\small f}
    \put(1, 63) {\small g}
    \put(1, 52) {\small h}
    \put(1, 39) {\small i}
    \put(1, 29.5) {\small j}
    \put(1, 18) {\small k}
    \put(1, 8.2) {\small l}
    \end{overpic}
    \caption{{\bf Result gallery.} Various modeling sequences created during design sessions using Sketch2CAD. The corresponding protocol steps are shown in the supplemental material. Please also refer to the supplementary video. }
    \label{fig:result_gallery}
\end{figure*}
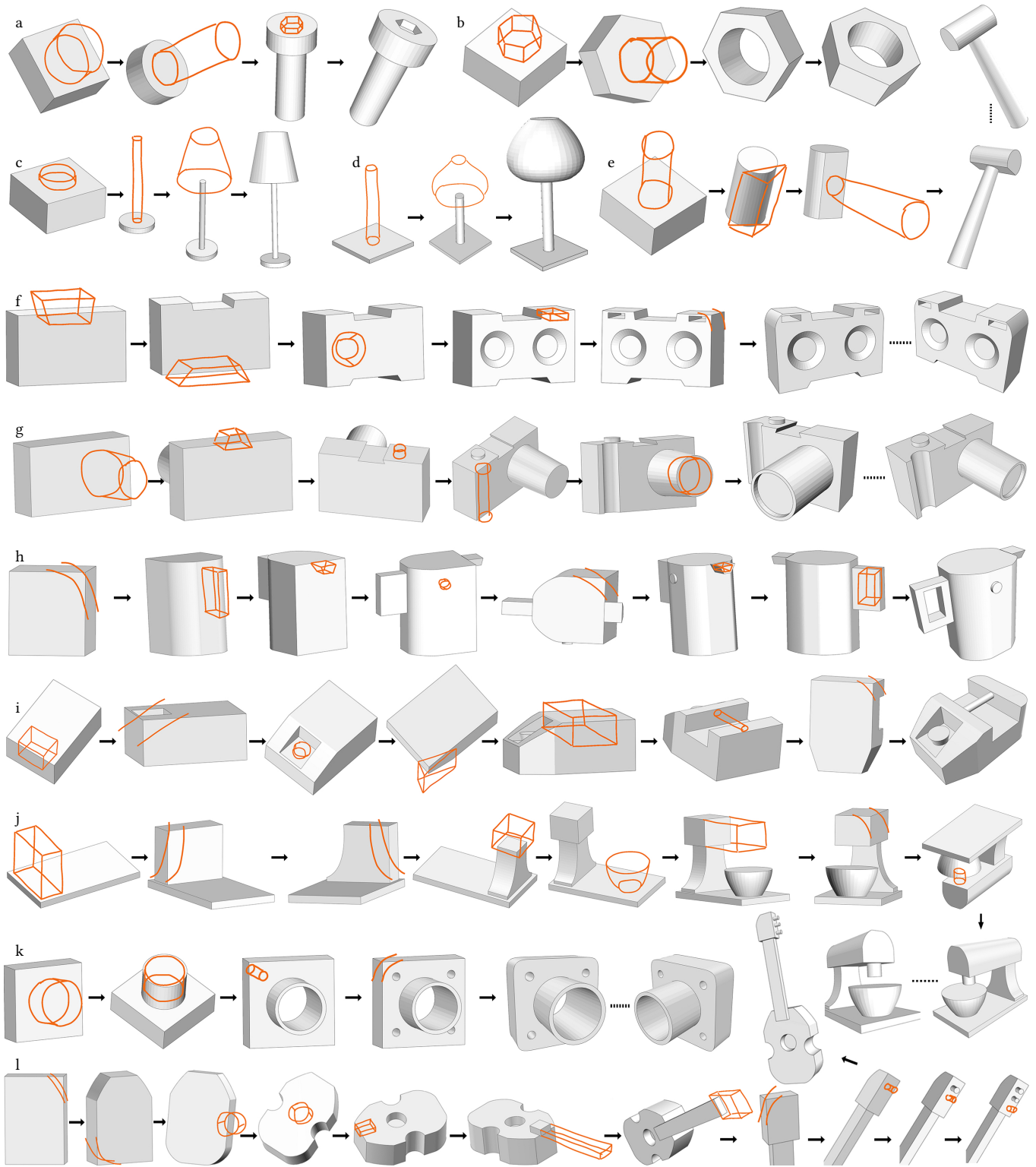

\paragraph{Regularization.}
In addition to the inherent regularization enabled by casting sketch into predefined operations in the procedural language level, 
our interface applies curve-level regularization, like snapping and rectification, to assist user sketching, as is commonly found in CAD modeling software.
The general idea of snapping is to detect key points, e.g., centers, edge middle points, corners of the base face, and align the corners and centers of the sketched primitive shape with them, whenever the point pair comes within a (default) small distance.
The general idea of rectification is to detect the approximate parallelism between sketched edges and base face edges, as well as the approximate equality of corner angles/edge lengths of sketched $N$-gons, and enforce the parallelism and equality by constructing parallel edges and regular $N$-gons analytically.

In particular, the snapping happens when the distance between the nearest key point pairs is within $10\%$ of the diameter of the base face.
The rectification happens when the differences of edge angles from zero, or of corner angles from $\frac{(N-2)180^\circ}{N}$, are within 20$^\circ$, for parallelism and corner equalization, respectively.
It also happens when the differences of side lengths are within $20\%$ of the average length for side length equalization.
An example of regularization for rectangular prism addition is shown in Fig.~\ref{fig:regularization}.
The user can switch off the auto-regularization to sketch arbitrary shapes.

\paragraph{Tuning operation parameters.} As an advantage of inferring CAD operations the users can edit the recovered parameters of a sketched operation. We support three types of adjustments:  creation of base shapes, resolution of ambiguous results, and fine tuning for geometric precision.
First, users can select Boolean operations between  existing shape and the sketched primitive, which is useful for quickly creating a base shape different from the plain box that the system starts with. 
Second, users can switch between the union and difference options for \texttt{add/subtract} and \texttt{sweep}, which have inherent ambiguous cases that require user specification (Sec.~\ref{sec:networks}).
Third, users can fine-tune the geometric parameters, e.g., offset distances, circle radius, etc.
In particular for a swept shape, when tuning the distance between circles or the radius of a circle, we adjust the control points of the profile cubic Bezier curve in proportion, defined by the distance from a control point to the fixed base circle or rotational axis, to preserve the overall shape of the swept geometry as much as possible.
An example of editing parameters of sequential operations is shown in Fig.~\ref{fig:parameter_tuning}.

\paragraph{Auto-completion by symmetry.}
Symmetry is prevalent in CAD models and can greatly ease user interaction.
In our tool, the user can take advantage of symmetry by reflecting a sketched primitive shape \cite{peng2018autocomplete} and its Boolean operation through the selected cross section planes of the oriented bounding box of the base shape,
thus avoiding the need to repeat the sketch multiple times manually.
An example of auto-completion by symmetry is shown in Fig.~\ref{fig:regularization}.
More examples are shown in Figs.~\ref{fig:teaser} and \ref{fig:result_gallery}.

\section{Results and Discussion}

With our tool, we have sketched several models of different complexities.
Examples are shown in Figs.~\ref{fig:teaser}  and~\ref{fig:result_gallery}, with the operation sequences ranging from 2 to \rev{11} steps, constructing CAD models from the simple bolts and nuts to the sophisticated mixer and cameras.
User evaluation also confirms the ease of sketching CAD models with our approach (Sec.~\ref{sec:user_study}).
We also validate the important design choices in our framework through ablation tests in Sec.~\ref{sec:ablation},
and discuss limitations and future work in Sec.~\ref{sec:future_work}.
Interactive modeling sessions, complete user evaluation data, model mesh files and sample protocol files can be found in the supplemental material.

\paragraph{Runtime}
\rev{Tested on a desktop PC with Intel(R) Core i9-9900 3.1GHz CPU and NVidia RTX 2070 Super GPU, the network inference is instantaneous, taking around 0.07s. 
Most time is spent on line search, ranging from 0.01s to 2s, as the step involves repeated computation of distances between pixels and stroke points, although it can be largely parallelized. 
To apply the recovered operator, a Boolean mesh operation typically takes around 0.02s.
}

\subsection{User evaluation}
\label{sec:user_study}

\begin{figure}[t!]
    \centering
    \begin{overpic}[width=\linewidth]{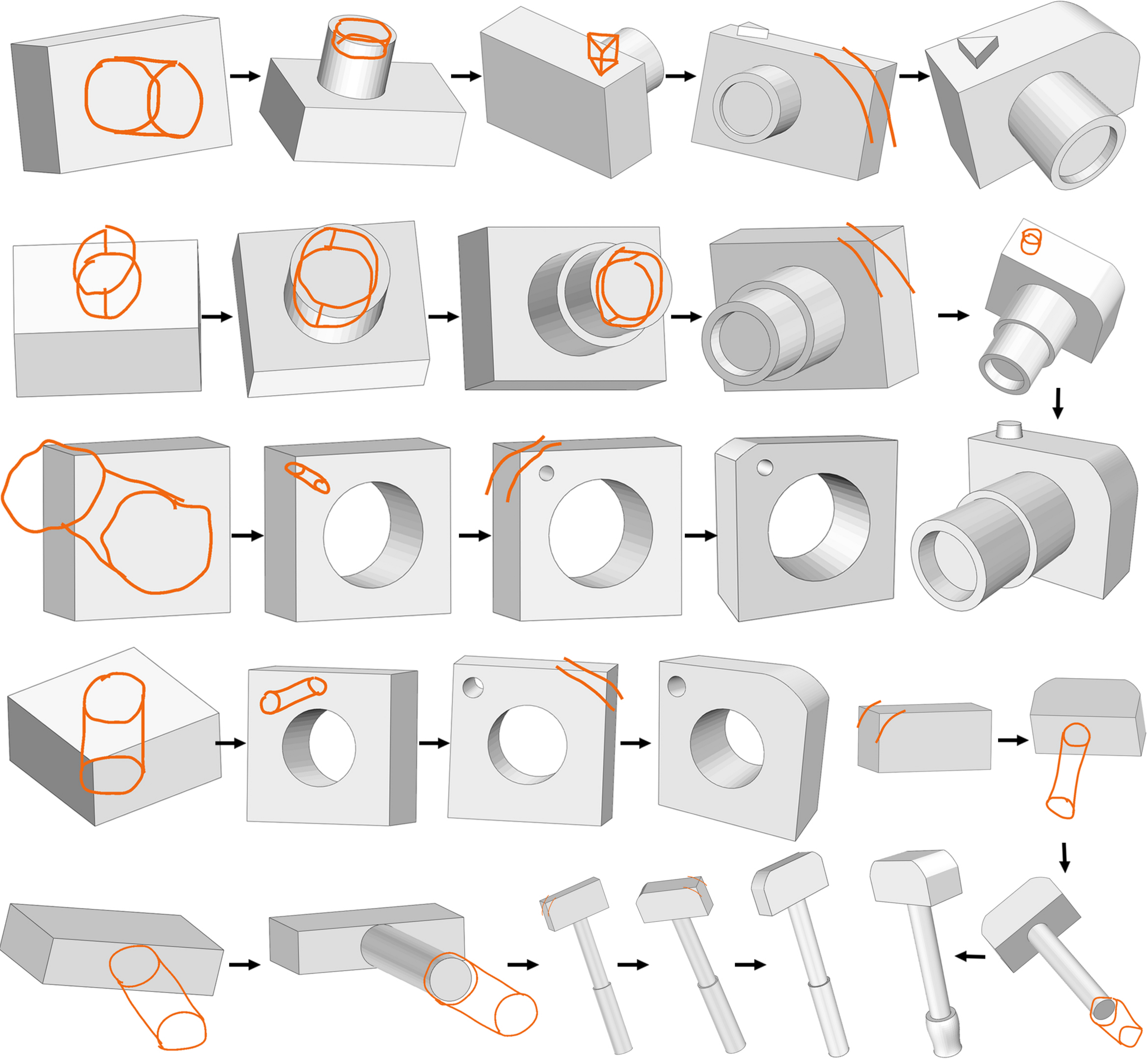}
     \end{overpic}
    \caption{{\bf Selected user modeling steps.} Users envision different paths and variations of operations for reaching the similar targets. The freehand inaccurate sketches are robustly translated into intended operations.}
    \label{fig:user_steps}
    \vspace{-4mm}
\end{figure}

We have evaluated the ease of use of our system by asking $6$ novices~\footnote{Due to requirement for GPU at inference time and restriction on lab access, we could not test the system with a wider set of users.} to create the same $3$ reference shapes. 
The target shapes, pre-modeled by an expert user, were presented to participants as a static image (Figure~\ref{fig:user_gallery}, green). 
\rev{Nevertheless, we do allow the participants to try and explore with variations, so that novel and interesting deviations from the references can be expected.}
All participants had little to no experience in sketching nor in CAD modeling, and were given a tutorial and a short practice session to get familiar with our system. 

Figure~\ref{fig:user_gallery} shows all models created by the participants, along with their time to completion. On the one hand, all participants managed to quickly produce models that closely match the reference shapes, demonstrating the ability of our system to make CAD modeling accessible to non-professionals. On the other hand, several participants also decided to deviate from the reference, for instance by adding a second part to the lens of the camera (P2), or by modeling a curved handle for the hammer (P5). We see this unexpected behavior as a consequence of the joy and artistic freedom offered by freehand sketching. 

Figure~\ref{fig:user_steps} provides a selection of intermediate sketching steps performed by the participants, which shows that our system is capable of interpreting a wide variety of strokes representing similar shapes. Note that all participants used a mouse to draw the input strokes, which our system nevertheless translates into 
\rev{regularized} CAD operations. Several participants commented that they appreciated the ability of our system to produce regular shapes from approximate strokes, and that they prefer to let the modeling flow going rather than revise what they had drawn. Although given the option, none of the user turned off the stroke regularizer option in Sketch2CAD. 
\rev{Participants gave an average rating of 4.75 on a 5-point Likert scale when asked whether the sketches are properly translated to CAD operations, and an average rating of 5 for ease of conception of the modeling sequences. Complete user feedback and comments on the ease of use of both the sequential modeling paradigm and our prototype tool can be found in the supplemental material.}

\subsection{Ablation study}
\label{sec:ablation}

\begin{figure}[b!]
    \centering
    \includegraphics[width=0.9\linewidth]{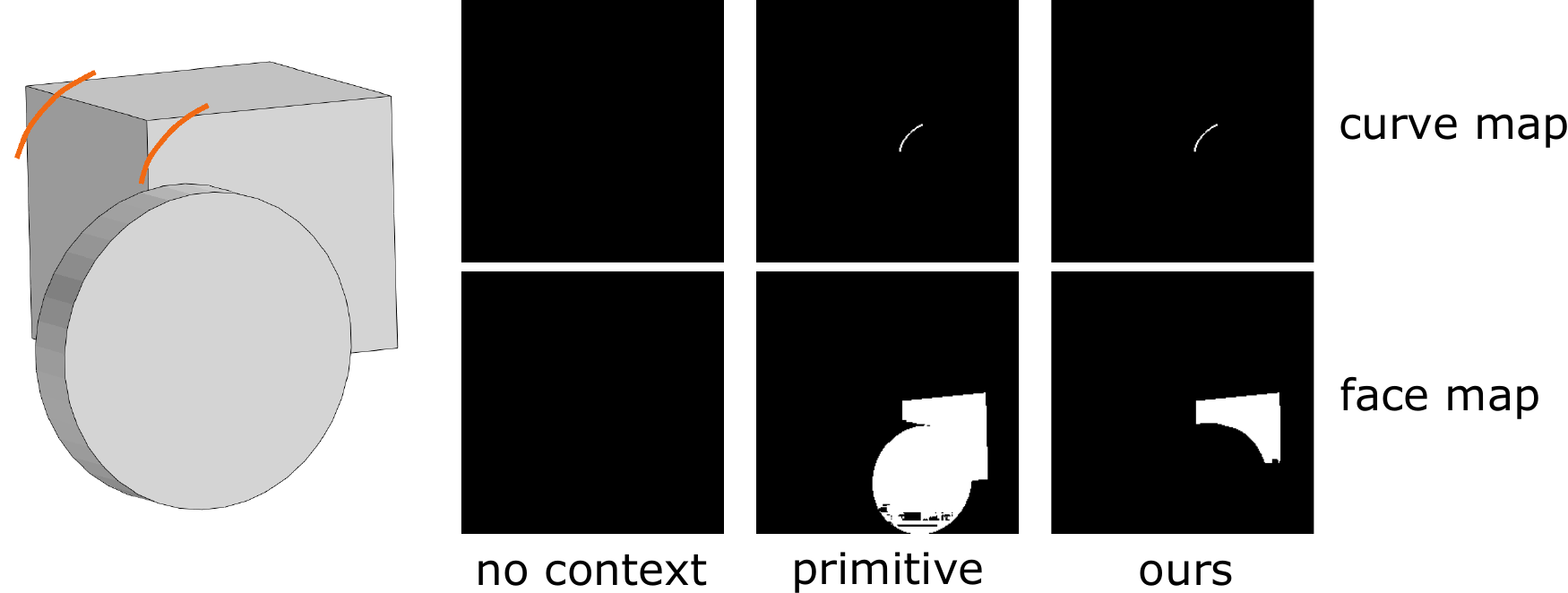} 
    \vspace{-3mm}
    \caption{\textbf{Comparing the ablation configurations by example.} The real bevel sketch is shown on the left. The `no context' network fails to produce any output that is above the map threshold (Sec.~\ref{sec:networks}). The `primitive' network predicts a good curve map, but cannot distinguish the two front facing polygons for base face map, which leads to the wrong base face detected by counting (Sec.~\ref{sec:networks}). Our full network gives almost perfect base face and curve segmentation maps. }
    \label{fig:ablation_examples}
\end{figure}

We validate two key components of our framework by ablation tests evaluated on the segmentation tasks for all operations:
\begin{enumerate}
    \item Using \textit{local context} versus using sketch only. The comparison is shown in Table~\ref{tab:context_prim_ablation}, where the `no context' configuration uses only the sketch map as network input.
    It is clear that without the local context maps of depth and normal of existing shapes, the segmentation of both the base face and the sketch curves becomes very difficult, with the base face IoU frequently under 10\%.
    In real user sketching, the networks without context are barely usable (Fig.~\ref{fig:ablation_examples}).
    \item Using shapes composed by multiple operations for network training, versus using primitive shapes only.
    The configuration of `primitive' shown in Table~\ref{tab:context_prim_ablation} trains the regression networks on another set of 40k synthesized shapes and sketches, which however only contains sequences of length 1 (Sec.~\ref{sec:network_training}).
    The `primitive' networks are then evaluated on the same 10k testing dataset of different sequence lengths (Sec.~\ref{sec:network_training}) and compared with our results.
    It is clear that the primitive networks that do not see sufficiently complex combinations of operation cannot match the accuracy of our results, with base face IoU lower for more than 10\%. 
    Real tests by user sketching show the difference as well (Fig.~\ref{fig:ablation_examples}).
\end{enumerate}
In addition, we note that for the operator classification task, since the four operations have quite different sketch patterns, the two ablated configurations can achieve comparable performances as our full network does, \rev{i.e., $99.80\%$ of no context, $93.68\%$ of primitive and $99.79\%$ of ours, since depth and normal maps do not play the essential role in operation classification.}

\begin{table}[t!]
\caption{Ablation tests on using context as input and using operator composition to generate training data. Our full network using context as input and trained on synthesized shapes composed by multiple operators has the best accuracy for all segmentation tasks.}
\vspace{-3mm}
\label{tab:context_prim_ablation}
  \begin{tabular}{r|c|c|c} 
      \hline
      Operator & Config & Face IoU(\%) & Curve Acc. (\%) \\
      \hline\hline
      \multirow{3}{*}{\texttt{Extrude}} & no context & 59.31 & 98.20 \\
       & primitive & 70.27 & 87.78 \\
       & ours & \textbf{91.78} & \textbf{98.04} \\
      \hline
      \multirow{3}{*}{\texttt{Bevel}} & no context & 3.23 & 50.24 \\
       & primitive & 73.27 & 93.88 \\
       & ours & \textbf{91.73} & \textbf{98.12} \\
      \hline 
      \multirow{3}{*}{\texttt{Add/Sub}} & no context & 10.27 & 66.21 \\
       & primitive & 63.81 & 87.32 \\
       & ours & \textbf{78.26} & \textbf{93.54} \\
      \hline
      \multirow{3}{*}{\texttt{Sweep}} & no context & 7.34 & 48.90 \\
       & primitive & 68.72 & 90.10 \\
       & ours & \textbf{78.34} & \textbf{95.57} \\
      \hline 
  \end{tabular}
  \vspace{-4mm}
\end{table}

\subsection{Robustness of network predictions}

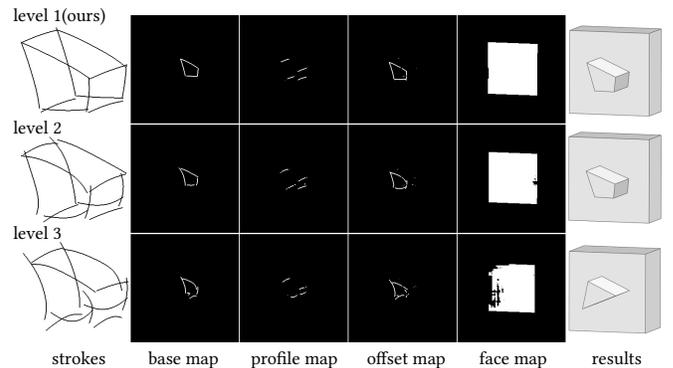
\begin{figure}[b!]
    \centering
    \begin{overpic}[width=\linewidth]{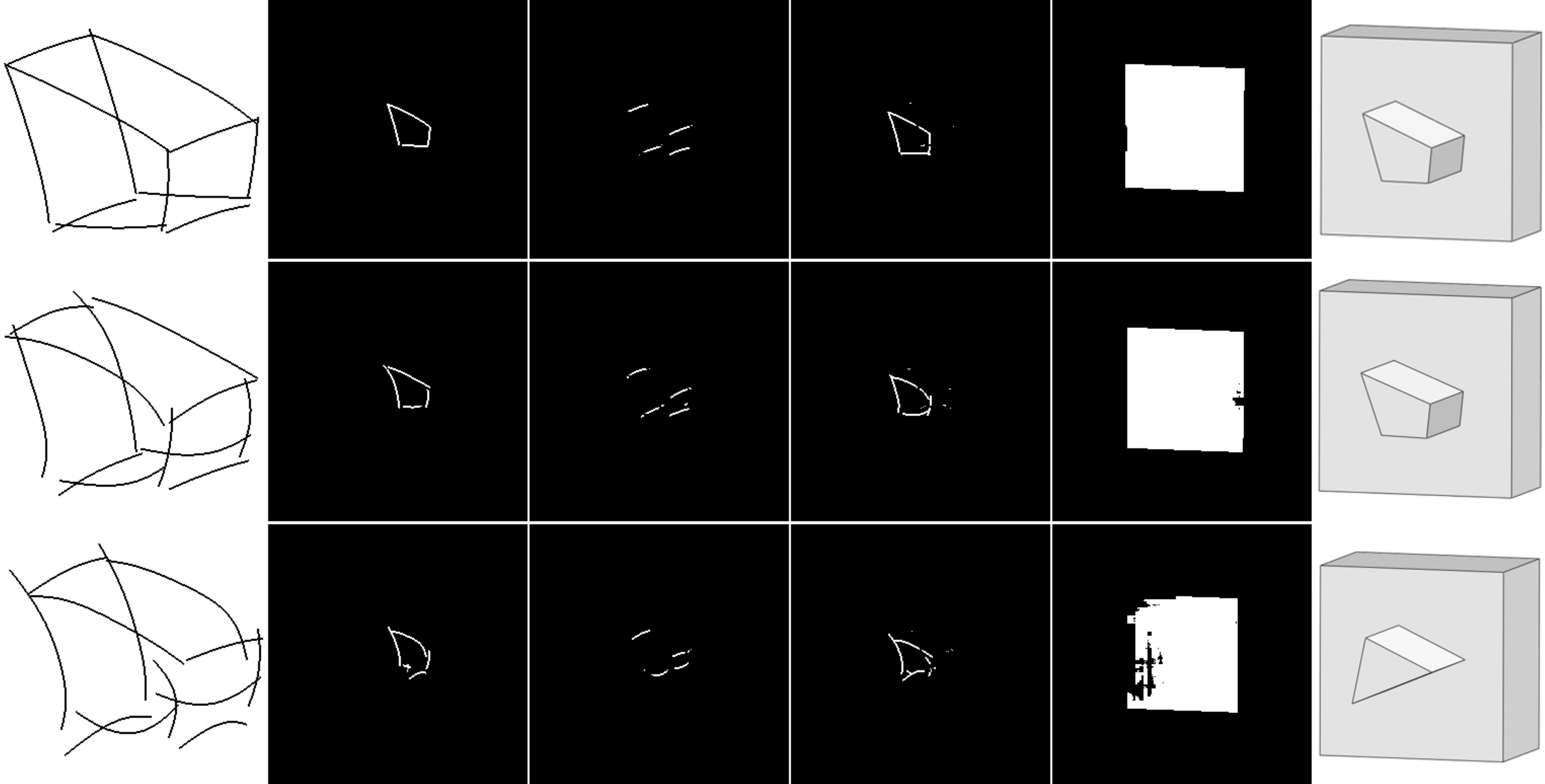}
        \put(-1, 50) {\footnotesize level 1(ours)}
        \put(-1, 32.5) {\footnotesize level 2}
        \put(-1, 16) {\footnotesize level 3}
        \put(5, -3.5) {\footnotesize strokes}
        \put(20, -3.5) {\footnotesize base map}
        \put(36, -3.5) {\footnotesize profile map}
        \put(54, -3.5) {\footnotesize offset map}
        \put(71.5, -3.5) {\footnotesize face map}
        \put(89, -3.5) {\footnotesize results}
    \end{overpic}
    \caption{\rev{\textbf{Sketch perturbation examples and the corresponding network predictions and fitting results.} Each row shows an example of a specific stroke perturbation level, while different columns show strokes, network outputs and the final result after the parameter fitting. }}
    \label{fig:noise_example}
\end{figure}

We test the robustness of network predictions under increasing levels of sketch irregularity.
While it is difficult to collect large amounts of real user sketches with different levels of irregularity, we simulate the variations by adding perturbations to clean sketches, as done for the synthetic training data generation (Sec.~\ref{sec:network_training}).
To be specific, we add stronger stroke perturbations than the training data generation configuration, i.e., $1.4\%$ of the rendered image diagonal length for level 1(ours), $2.8\%$ for level 2 and $4.1\%$ for level 3 (see Fig. \ref{fig:noise_example}), and evaluate how the pretrained model works under such out-of-distribution settings.
The statistics are reported in Table \ref{tab:noise_ablation} and example sketches are shown in Fig. \ref{fig:noise_example}. Quantitatively, as the noise increases, the segmentation networks produce more inaccurate results, and the same observation is found from the classification network ($99.79\%$ of level 1~(ours), $85.06\%$ of level 2, and $53.20\%$ of level 3, respectively). 
Qualitatively, while the segmentation network was trained with a low level of noise, it produces high-quality segmentation maps for moderate noise (level 2). While high noise (level 3) degrades the segmentation maps, the subsequent parameter fitting still yields a reasonable shape.

\begin{table}[t!]
\caption{\rev{Quantitative robustness test of network predictions. Both the face IoU and curve regression accuracy drop noticeably as the noise increase.}}
\vspace{-3mm}
\label{tab:noise_ablation}
  \begin{tabular}{r|c|c|c}
      \hline
      Operator & Config & Face IoU(\%) & Curve Acc. (\%) \\
      \hline\hline
      \multirow{3}{*}{\texttt{Extrude}} 
       & level 1(ours) & \textbf{91.78} & \textbf{98.04} \\
       & level 2 & 89.59 & 96.28 \\
       & level 3 & 82.58 & 91.69 \\
      \hline
      \multirow{3}{*}{\texttt{Bevel}} 
       & level 1(ours) & \textbf{91.73} & \textbf{98.12} \\
       & level 2 & 85.90 & 94.89 \\
       & level 3 & 75.76 & 90.27 \\
      \hline 
      \multirow{3}{*}{\texttt{Add/Sub}} 
       & level 1(ours) & \textbf{78.26} & \textbf{93.54} \\
       & level 2 & 77.09 & 92.02 \\
       & level 3 & 73.88 & 86.08 \\
      \hline
      \multirow{3}{*}{\texttt{Sweep}} 
       & level 1(ours) & \textbf{78.34} & \textbf{95.57} \\
       & level 2 & 77.31 & 94.44 \\
       & level 3 & 75.47 & 92.78 \\
      \hline 
  \end{tabular}
  \vspace{-4.5mm}
\end{table}

\subsection{Limitations and Future Work}
\label{sec:future_work}

In its current form, Sketch2CAD does not support drawing primitives on curved faces (e.g., on the curved face of a cylinder). One possibility would be to use NURBS as the modeling primitives, where stitching face can be curved NURBS patches stopping at trim lines. This would, however, require an extension of the underlying geometry engine used in our implementation. 
Another limitation involves drawing small features (e.g., knobs, or screw threads). While we do support zoom in our interface, having a library of small leaf-level part features can be useful to instantiate, rather than build up from scratch. Finally, we expect a certain amount of sketching ability from the user. Porting our code to a tablet interface can further lower this entry bar.

\begin{figure}[t!]
    \centering
    \includegraphics[width=0.9\linewidth]{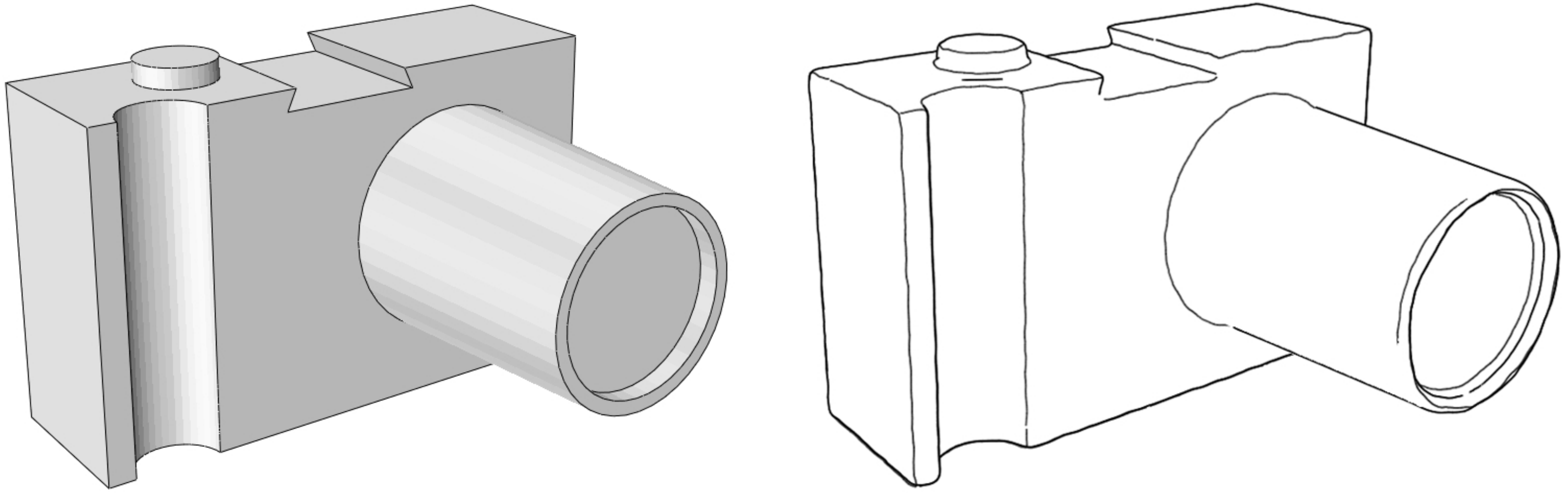} 
    \caption{\textbf{CAD-based sketching.} The CAD models generated in our system can subsequently be procedurally edited -- subdivided and smoothed in this example -- and the resultant mesh be used to go back to a `sketch' using NPR rendering. This allows the user to perform operations that are much easier in the CAD domain and then transition back to sketching, possibly with camera view changes. This can be useful during ideation and prototyping phases of product design.  }
    \label{fig:cad2sketch}
    \vspace{-2mm}
\end{figure}

Our training data generation only considers geometric feasibility rather than semantics, e.g., not all combinations of the operations are functionally meaningful. 
On the other hand, in the scenario of CAD modeling, there are strong semantics about the desired forms and functions of the different parts and their composing operations for common man-made objects.
In the future, we plan to take this factor into consideration and train our networks on more realistic data that respect real world model distributions, e.g.,  by utilizing dataset with semantic annotations like PartNet~\cite{PartNet}. It will also be interesting to train our network directly on CAD modeling trace data, when available, to capture typical sequences of operations and learn auto-complete routines (cf., \cite{peng2018autocomplete}) directly from user data. Finally, while in this work we explored sketch-to-CAD, we can easily use the generated models, possibly after 3D based editing and manipulations, to go back to the sketch domain and thus enable powerful edits to sketching. Figure~\ref{fig:cad2sketch} shows an early example of such a possible workflow. 

\section{Conclusion}
The visceral and approximate nature of freehand sketching is often considered to be in contradiction with the tediousness and rigidity of 3D modeling. Yet, we observed that industrial design sketching and CAD modeling follow very similar workflows, where practitioners create complex shapes as a sequence of simple sketching (resp. modeling) operations. By identifying and parameterizing common operations in the two domains, and training a deep neural network to recognize and segment these operations, we offer an interactive system capable of turning approximate sketches of \rev{human-made} objects into \rev{regular} CAD models, \rev{as illustrated by our} evaluation with novices as well as the diversity of shapes we created with our approach.

\begin{acks}
\rev{The authors would like to thank the reviewers for their valuable and detailed suggestions, the user evaluation participants and Nathan Carr, Yuxiao Guo, Zhiming Cui for the valuable discussions. The work of Niloy was supported by ERC Grant (SmartGeometry 335373), Google Faculty Award and gifts from Adobe, and the work of Adrien was supported by ERC Starting Grant D3 (ERC-2016-STG 714221), research and software donations from Adobe. Finally, Changjian Li wants to thank, in particular, the endless and invaluable love and supports from Huahua Guo over the tough time due to COVID-19.}
\end{acks}

\bibliographystyle{ACM-Reference-Format}
\bibliography{src/sketch2CADBib}

\end{document}